\documentclass{JHEP3}
\usepackage{graphicx,epsfig}
\usepackage{cite}
\usepackage[english]{babel}
\usepackage{amssymb,amsfonts,amsmath}
\usepackage{verbatim}
  \newcommand{\cH}{{\cal H}}

  \newcommand{\cN}{{\cal N}}
  \newcommand{\cP}{{\cal P}}
  
\newcommand{\cS}{{\cal S}}

\newcommand{\be}{\begin{equation}} \newcommand{\ee}{\end{equation}}
\newcommand{\bea}{\begin{eqnarray}} \newcommand{\eea}{\end{eqnarray}}
\newcommand{\beann}{\begin{eqnarray*}}  \newcommand{\eeann}{\end{eqnarray*}}
\newcommand{\bfig}{\begin{figure}} \newcommand{\efig}{\end{figure}}
\newcommand{\ba}{\begin{array}} \newcommand{\ea}{\end{array}}
\newcommand{\bcen}{\begin{center}} \newcommand{\ecen}{\end{center}}
\newcommand{\btab}{\begin{tabular}} \newcommand{\etab}{\end{tabular}}

\def\tr{\operatorname{tr\:}}

\newcommand{\bra}[1]{\langle #1|}
\newcommand{\ket}[1]{|#1\rangle}
\newcommand{\vev}[1]{\left\langle{#1}\right\rangle}

\newtheorem{Proposition}{Proposition}[section]

\newtheorem{Theorem}{Theorem}[section]
\newtheorem{Lemma}{Lemma}[section]
\newtheorem{Corrolary}{Corrolary}[section]

\newcommand{\bp}{\begin{Proposition}}   \newcommand{\ep}{\end{Proposition}}
\newcommand{\bt}{\begin{Theorem}}   \newcommand{\et}{\end{Theorem}}
\newcommand{\bl}{\begin{Lemma}}     \newcommand{\el}{\end{Lemma}}
\newcommand{\bc}{\begin{Corrolary}} \newcommand{\ec}{\end{Corrolary}}

\def\half{\frac{1}{2}}
\def\bb#1{{\pmb{#1}}}

\title{Ward Identities for Hall Transport}

\author{Carlos Hoyos, Bom Soo Kim, Yaron Oz\\
Raymond and Beverly Sackler School of
Physics and Astronomy, Tel-Aviv University, Tel-Aviv 69978, Israel\\
E-mail: \email{choyos,bskim,yaronoz@post.tau.ac.il}}

\abstract{We derive quantum field theory Ward identities based on linear area preserving and conformal transformations in 2+1 dimensions.
The identities relate Hall viscosities, Hall conductivities and the angular momentum. They apply both for relativistic
and non relativistic systems, at zero and at finite temperature.
We consider systems with or without translation invariance, and introduce an external magnetic field and
viscous drag terms.
A special case of the identities yields the well known relation between the Hall conductivity and half the angular
momentum density.

}
\preprint{TAUP-2984/14}

\begin{document}


\section{Introduction}

Quantum field theory Ward identities are relations among correlators that are derived using the symmetry generators.
They are valuable even when the symmetries are broken.  Of particular
interest are Ward identities associated with space-time transformations. In this paper
we will derive quantum field theory Ward identities based on linear area preserving and conformal transformations.
The identities yield relations among different quantities such as viscosities, conductivities and angular momentum
of the system.

We will consider parity breaking quantum field theories  in $2+1$ space-time dimensions. There is a large number of such
systems of interest, for instance the Quantum Hall states.
They exhibit non-dissipative parity breaking transport properties, such as the well known  Hall viscosity $\eta_H$.
The Hall viscosity in $2+1$ dimensional quantum systems was studied first in \cite{Avron:1995},\footnote{Similar transport coefficients in classical magnetized plasmas were discussed before \cite{Landau10}.} and has been much studied since, see \cite{Avron:1997,Hoyos:2014pba} for reviews. Its value in topological states such as Hall states or chiral superfluids has been computed in many different systems
\cite{Avron:1995,Levay1995,Tokatly2007,Read:2008rn,Tokatly2009,Read2011,Bradlyn:2012ea,Cho2014}, and it has been shown that its value divided by the particle number density is quantized. This makes it particularly interesting, since it gives a characterization of the state independent of the Hall conductivity. A closely related quantity is the torsional Hall viscosity that was introduced for relativistic theories in
\cite{Hughes:2011hv,Hughes:2012vg}.

From an effective field theory point of view a non-zero Hall viscosity is expected on general grounds in theories with broken parity
\cite{Nicolis:2011ey,Hoyos:2011ez,Hidaka:2013,Hoyos:2013eha,Haehl:2013kra,Geracie:2014iva}, including hydrodynamics \cite{Jensen:2011xb,Kaminski2013}.
Hall viscosity was also introduced in holographic models in \cite{Saremi:2011ab,Son:2013xra}, and further studied in
\cite{Chen2011,Chen2012,Cai:2012mg,Liu:2014gto,Hoyos:2014nua}.

A novel formula
relating the Hall viscosity to the angular momentum density of the system $\ell$
\begin{equation}
\eta_H=\ell/2 \ ,
\label{novel}
\end{equation}
 was
shown to hold  in certain non-relativistic systems on the torus
in \cite{Read:2008rn,Read2011}.
The relation between the Hall conductivity and the angular
momentum density has been studied in various setups \cite{Nicolis:2011ey,Son:2013xra,Liu:2014gto,Hoyos:2014nua}. In \cite{Son:2013xra} it was argued that the relation between angular momentum and Hall viscosity holds in a holographic chiral superfluid. However, angular momentum can be introduced in several ways (see \cite{Liu:2012zm,Wu:2013vya}) and generically it is not simply related to the Hall viscosity \cite{Liu:2014gto}, which could even vanish. These results
are based on bottom-up models.  It is of interest to have a general field theory argument that clarifies the relation between the angular momentum and the Hall viscosity, and that is also applicable to relativistic theories. This is one of the motivations for the present work.

In \cite{Bradlyn:2012ea} it has been shown that Ward identities of non-relativistic systems
lead to a non-trivial relation between Hall viscosity and conductivities, and between Hall viscosity and angular momentum density.
In this paper we will derive such identities for general relativistic or non-relativistic $2+1$ dimensional quantum field theories,
at zero and at finite temperature.
We will consider systems with or without translation invariance, and introduce an external magnetic field and viscous drag terms.
In particular we will see under what conditions does the relation (\ref{novel}) hold.

The paper is organized as follows.
In section \ref{sec:ward} we will introduce the algebra of linear area preserving symmetries and derive the corresponding Ward
identities in general systems. Having angular momentum density breaks translation invariance and the Ward identity takes the form
(\ref{rel2}). We will then introduce an external magnetic field and obtain the relation (\ref{I13solB})
with the modified angular momentum (\ref{lB}). In the presence of viscous drag terms, that are of relevance for instance to systems
with a lattice or impurities, we will derive the relation (\ref{I13solBdis}).
In section \ref{sec:cond} we will consider the relation between the Hall viscosity and conductivities in translationally invariant systems. In particular, we will derive the
relation (\ref{etakappa}) and with viscous terms (\ref{etahfin}).
We will show that these formulas reduce in the non-relativistic limit and in the absence of drag terms to
the one derived in \cite{Bradlyn:2012ea}. The latter has been verified explicitly in some models.

In section \ref{sec:ang} we will derive the Ward identity relating the Hall viscosity to the angular momentum density when translation invariance is broken.
We will obtain (\ref{etal}) and (\ref{simpler}) that in particular
for gapped system yield  (\ref{novel}). We will argue that in the limit where translation invariance is recovered the value of the Hall viscosity does not change.
In section \ref{sec:conf} we will consider spatial conformal transformations on the plane and use them in order to derive the Ward identity relation (\ref{conformal}), which we use in order to show the relation between the angular momentum density and the total
pressure of the system including the Hall bulk viscosity contribution (\ref{pressurerelation}).
In the appendices we provide details of the calculations as well as a generalization to nonzero temperature.

For convenience we introduced in table~\ref{table:definitions} all the notations that will be used in the paper, and highlighted
the main results in the different sections.

\TABLE[hb]{
\caption{Definitions and notations}
\centering
\begin{tabular}{l c l}
\hline\hline
$t, \hat t$ & &  Time variables with Fourier frequency $\omega$   \\
$\bb{x}, \bb{\hat{x}}$ & & Space variables with Fourier momentum $\bb{p}, \bb{q}$ \\
$\bb{P}=\frac{\bb{p}+\bb{q}}{2},~ \bb{k}=\bb{p}-\bb{q}$ & &  \\
$S_a^\mu$ & & Area preserving current density, $a=1,2,3$ \\
$Q_a = \int d^2 \bb{x} ~S_a^0$ & & Three corresponding charges  \\
$\bar \sigma_1= \sigma_1, \bar\sigma_2=i\sigma_2, \bar \sigma_3=\sigma_3$  & & $\sigma_a$ : Pauli matrices  \\
$S_{ijkl}=\frac{1}{4}(\bar \sigma_1)_{ij}(\bar \sigma_3)_{kl}$  & &   \\
$\tilde \ell, ~\bar \ell$ & & Angular momentum density, its average   \\
$\tilde k,~ \bar k$ & & Generalized $\tilde \ell, \bar \ell$ for $B\neq 0$  \\
$\langle L_{xy}\rangle$ & & Total angular momentum  \\
$M(\bb{x})$ & & Space dependent magnetization  \\
$P_0$ & & Constant pressure  \\
$\bb{\epsilon}$ & & Regulator \\
$\delta$ & & Scale of breaking of translation invariance \\
$\eta_H, ~\bar \eta_H$ & & Hall viscosity, its average  \\
$\sigma^{ij}, ~\kappa^{ij},~\alpha^{ij}, ~\bar \alpha^{ij} $ & & Electric, momentum, mixed  conductivities \\
$\sigma_H = \frac{1}{2} \epsilon_{ij}\sigma^{ij}, ~\kappa_H = \frac{1}{2} \epsilon_{ij} \kappa^{ij}$ & & Hall and Hall momentum conductivities \\
$\text{tr} ~\sigma= \delta_{ij}\sigma^{ij} $ & &  Trace of electric conductivity\\
$\text{tr} ~\alpha= \delta_{ij}\alpha^{ij}, ~\text{tr} ~\bar \alpha= \delta_{ij}\bar \alpha^{ij} $ & &  Trace of mixed momentum-current conductivities\\
$\bar n$ & & Charge density \\
$\omega_c= \frac{B}{m}$ & & Cyclotron frequency \\
$\lambda_J, ~\lambda_T, ~\lambda_{NR}=\frac{\lambda_J}{m}+\lambda_T $ & & Drag coefficients  \\
$V_1^i=T^{0i}, ~V_2^i=\epsilon^i_nJ^n$ & & $T$: energy-momentum tensor, $J$: current \\
\hline\hline
\end{tabular}
\label{table:definitions}
}

\section{Ward identities from linear transformations}
\label{sec:ward}

Ward identities are constraints among correlators that can be derived by taking advantage of the symmetries of the theory. They are useful even when the would-be symmetries are broken. In particular, when the symmetries are related to spacetime transformations, there are various relations that are of much
 physical interest.  Canonical examples are conformal transformations, where Ward identities provide a deep insight about the properties of the theory.

The generators of linear transformations in $d+1$ dimensions are
\begin{equation}
Q^{\mu\nu} = \int d^d \bb{x}\, x^{\mu} T^{0\nu},
\end{equation}
where $T^{\mu\nu}$ is the energy-momentum tensor (We will use Greek indices $\mu,\nu=0,1,\cdots,d$ for spacetime directions and Latin indices $i,j=1,2,\cdots,d$ for purely spatial directions). If Lorentz invariance is broken then it may be of interest to study the generators of spatial transformations $Q^{ij}$. An interesting subgroup comprises the area-preserving linear transformations. It is quite natural to consider them for instance in Quantum Hall systems, where the effective description is an incompressible fluid.

As was shown in \cite{Bradlyn:2012ea} for non-relativistic systems, the Ward identities lead to a non-trivial relation between Hall viscosity and conductivities, and between Hall viscosity and angular momentum density. In the following we will extend the analysis to generic quantum field theories in 2+1 dimensions, both relativistic and non-relativistic.

\subsection{Area-preserving transformations in 2+1 dimensions}

In 2+1 dimensions there are three generators that we can construct contracting with the Pauli matrices\footnote{$\bar{\sigma}_a$ are the Pauli matrices for $a=1,3$ and $\bar{\sigma}_2=i\sigma_2$.}
\begin{equation}
Q_a=\frac{1}{2}(\bar{\sigma}_a)_{ij}Q^{ij}=\int d^2 \bb{x}\, S_a^0.
\end{equation}
Where the associated current density is
\begin{equation}
S^\mu_a=\frac{1}{2}(\bar{\sigma}_a)_{ij}x^i T^{\mu j}.
\end{equation}
This generates the algebra of $SL(2,\mathbb{R})$
\begin{equation}
[Q_a,Q_b]=if_{ab}^{\ \ c}Q_c.
\end{equation}
The structure constants are antisymmetric on the first pair of indices and
\begin{equation}
f_{13}^{\ \ 2}=f_{12}^{ \ \ 3}= f_{23}^{\ \ 1}=+1.
\end{equation}

In general the charges $Q_1$ and $Q_3$ are not conserved, since
\begin{equation}
\partial_\mu S^\mu_a =\frac{1}{2}(\bar{\sigma}_a)_{ij}T^{ij} \ \ \Rightarrow \ \ \partial_t Q_a =\frac{1}{2}\int d^2x\, (\bar{\sigma}_a)_{ij}T^{ij}.
\end{equation}
$Q_2$ is a symmetry because rotational symmetry implies $T^{ij}=T^{ji}$. It generates $SO(2)$ rotations in the plane. Note that all $Q_a$ can be accidentally conserved in isotropic states $\vev{T^{ij}}\propto \delta^{ij}$. A representation of the generators acting on local operators is ($\bb{x}=(x,y)$)
\begin{equation}
Q_1=-\frac{i}{2}(x\partial_y+y\partial_x), \ \ Q_2=-\frac{i}{2}(x\partial_y-y\partial_x),\ \ Q_3=-\frac{i}{2}(x\partial_x-y\partial_y).
\end{equation}

The generators should satisfy the symmetry algebra.
Therefore, the equal time commutator should satisfy the Ward identity
\begin{equation}\label{ward}
\vev{[S_a^0(t,\bb{x}), S_b^0(t,\bb{\hat{x}})]}=if_{ab}^{\ \ c}\vev{S^0_c(t,\bb{x})}\delta^{(2)}(\bb{x}-\bb{\hat{x}}).
\end{equation}
In particular
\begin{equation}
\vev{i[S_1^0(t,\bb{x}), S_3^0(t,\bb{\hat{x}})]}=-\vev{S^0_2(t,\bb{x})}\delta^{(2)}(\bb{x}-\bb{\hat{x}}).
\end{equation}
Note that
\begin{equation}
\int d^2x \int d^2 \hat{x} \vev{i[S_1^0(t,\bb{x}), S_3^0(t,\bb{\hat{x}})]}=\vev{i[Q_1,Q_3]}=-\frac{\vev{L_{xy}}}{2},
\end{equation}
where $\vev{L_{xy}}$ is the total angular momentum.

\subsubsection{Systems with conserved momentum}

If momentum is conserved $\partial_\mu T^{\mu i}=0$, we can use directly the generators constructed with $S_a^0$ to derive a Ward identity that relates the Hall viscosity to current correlators and angular momentum density. When momentum is not conserved the definition of the $SL(2,\mathbb{R})$ generators has to be modified, as we will discuss for systems with a background magnetic field in the next section.

With the commutator of the area-preserving currents we can construct the retarded correlator
\begin{equation}\label{g13}
G_{13}(t-\hat{t};\bb{x},\bb{\hat{x}})=i\Theta(t-\hat{t})\vev{[S_1^0(t,\bb{x}), S_3^0(\hat{t},\bb{\hat{x}})]},
\end{equation}
and similarly the retarded correlators of the energy-momentum tensor are defined as
\begin{equation}
 G_R^{\mu\nu,\alpha\beta}(t-\hat{t};\bb{x},\bb{\hat{x}})=i\Theta(t-\hat{t})\vev{[T^{\mu\nu}(t,\bb{x}),T^{\alpha\beta}(\hat{t},\bb{\hat{x}})]}.
\end{equation}
We are assuming time translation invariance but not necessarily space translation invariance.

We will compute the time derivatives of $G_{13}$, integrated over the spatial directions. In order to regulate the integrals we introduce a constant vector $\bb{\epsilon}$. Eventually we will take the limit $\bb{\epsilon}\to \bb{0}$\footnote{Alternatively, one can think of the calculation as taking the Fourier transform of the correlator and taking the zero momentum limit in a symmetric way.}
\begin{equation}\label{I13}
\begin{split}
&I_{13}(\bb{\epsilon})\equiv\partial_t\partial_{\hat{t}}\int d^2 \bb{x}\,\, d^2 \bb{\hat{x}} \, e^{-i\bb{\epsilon}\cdot\bb{x}-i\bb{\epsilon}\cdot\bb{\hat{x}}}  \,G_{13}(t-\hat{t};\bb{x},\bb{\hat{x}})\\
&=\partial_t\partial_{\hat{t}}\int d^2 \bb{x}\, d^2 \bb{\hat{x}}\, e^{-i\bb{\epsilon}\cdot\bb{x}-i\bb{\epsilon}\cdot\bb{\hat{x}}}   \,S_{ijkl}x^i\hat{x}^k G_R^{0j,0l}(t-\hat{t};\bb{x},\bb{\hat{x}}),
\end{split}
\end{equation}
Where we have defined the tensor
\begin{equation}
S_{ijkl}=\frac{1}{4}(\bar{\sigma}_1)_{ij}(\bar{\sigma}_3)_{kl}.
\end{equation}
We collect some useful algebraic relations in Eq.~\eqref{Srels}.

We will now do the Fourier transform with respect to time of the expression above
\begin{equation}\label{FI13}
\widetilde{I}_{13}(\bb{\epsilon})=\int d(t-\hat{t}) e^{-i\omega(t-\hat{t})} I_{13}(\bb{\epsilon}),
\end{equation}
and use the Fourier transform of the retarded correlators
\begin{equation}
G_R^{\mu\nu\alpha\beta}(t-\hat{t};\bb{x},\bb{\hat{x}})=\int \frac{dp_0 d^2 \bb{p} d^2 \bb{q}}{(2\pi)^5} e^{ip_0 (t-\hat{t})+i\bb{p}\cdot\bb{x}-i\bb{q}\cdot\bb{\hat{x}}}\tilde{G}_R^{\mu\nu\alpha\beta}(p^0,\bb{p},\bb{q}).
\end{equation}
After some manipulations (the details are in the Appendix~\ref{app:cons}), the Fourier transform can be written in the following form:
\begin{equation}\label{I13sol}
\begin{split}
&\tilde{I}_{13}(\bb{\epsilon})=\omega^2\,S_{ijkl}\, \frac{\partial}{\partial p_i}\frac{\partial}{\partial q_k}\tilde{G}_R^{0j0l}(\omega,\bb{p},\bb{q})\Bigg|_{\bb{p}=\bb{\epsilon},\bb{q}=-\bb{\epsilon}}.
\end{split}
\end{equation}
We will now use the conservation equations of the energy-momentum tensor to relate this result to the correlator of two stress tensors. In the end this will give us a formula for the Hall viscosity.

Starting with \eqref{g13} and taking the time derivatives explicitly we can also write \eqref{I13} as
\begin{equation}\label{I13b}
\begin{split}
I_{13}(\bb{\epsilon}) &=
\int d^2 \bb{x}\,\, d^2 \bb{\hat{x}} \, e^{-i\bb{\epsilon}\cdot\bb{x}-i\bb{\epsilon}\cdot\bb{\hat{x}}}\left[\delta'(t-\hat{t})\delta^{(2)}(\bb{x}-\bb{\hat{x}})\vev{S^0_2(\bb{x})}\right.\\
&\left.+S_{ijkl}x^i\hat{x}^k \partial_n\partial_{\hat{m}} G_R^{njml}(t-\hat{t};\bb{x},\bb{\hat{x}}) \right].
\end{split}
\end{equation}
Where we have used the Ward identity for the equal time commutator \eqref{ward} and we have assumed that there is a time-independent angular momentum density $\vev{S^0_2(\bb{x})}$. The result is (see the Appendix~\ref{app:cons})
\begin{equation}\label{I13solb}
\begin{split}
\tilde{I}_{13}(\bb{\epsilon}) &=i\omega\frac{\tilde{\ell}(2\bb{\epsilon})}{2}+ S_{ijkl}\frac{\partial}{\partial p_i}\frac{\partial}{\partial q_k}\left[ p_n q_m \tilde{G}_R^{njml}(\omega,\bb{p},\bb{q})\right]_{\bb{p}=\bb{\epsilon},\bb{q}=-\bb{\epsilon}}.\\
\end{split}
\end{equation}
Where we have defined the Fourier transform of the angular momentum density as
\begin{equation}
\tilde{\ell}(\bb{k})=2\int d^2 \bb{x}\,e^{-i\bb{k}\cdot\bb{x}} \, \vev{S^0_2(\bb{x})}.
\end{equation}
Note that $\tilde{\ell}(\bb{0})=\vev{L_{xy}}$ is the total angular momentum if it is finite. Equating \eqref{I13sol} and \eqref{I13solb} we obtain the relation
\begin{equation}\label{rel2}
\boxed{
\begin{aligned}
i\omega\frac{\tilde{\ell}(2\bb{\epsilon})}{2} + & S_{ijkl}\frac{\partial}{\partial p_i}\frac{\partial}{\partial q_k}\left[ p_n q_m \tilde{G}_R^{njml}(\omega,\bb{p},\bb{q})\right]_{\bb{p}=\bb{\epsilon},\bb{q}=-\bb{\epsilon}}\\
&=\omega^2\,S_{ijkl}\, \frac{\partial}{\partial p_i}\frac{\partial}{\partial q_k}\tilde{G}_R^{0j0l}(\omega,\bb{p},\bb{q})\Bigg|_{\bb{p}=\bb{\epsilon},\bb{q}=-\bb{\epsilon}}.
\end{aligned}}
\end{equation}
In the absence of angular momentum density the relation above follows simply from the conservation of the energy-momentum tensor in a translationally invariant system. This shows that if there is a non-zero total angular momentum translation invariance should be broken.

\subsubsection{Systems with a magnetic field}

The Hall viscosity was first computed in Quantum Hall systems, where a background magnetic field is turned on. It is then of interest to extend the analysis to this case.
In the presence of a magnetic field $B$, the conservation equation of the energy-momentum tensor is modified to
\begin{equation}
\partial_\mu T^{\mu i}=B\epsilon^i_{\ j} J^j,
\end{equation}
where $J^\mu$ is the electromagnetic current and we assume that the magnetic field is constant. Note that the angular momentum density as defined above is not conserved, since
\begin{equation}
\partial_\mu S_2^\mu = -\frac{B}{2} x_i J^i.
\end{equation}
It is possible to define shear and angular momentum operators that obey the same conservation equations at zero and non-zero magnetic field\footnote{We thank Moshe Goldstein for discussions on this point.}
\begin{equation}\label{angmom}
S_{B\,a}^\mu =\frac{(\bar{\sigma}_a)_{ij}}{2}x^i \left(T^{\mu j}-\frac{B}{2}\epsilon^j_{\ n}x^n J^\mu\right).
\end{equation}
The $SL(2,\mathbb{R})$ algebra is actually generated by these operators, rather than by the original $S_a^\mu$. We will discuss this in more detail when we present the relation to angular momentum density.

Even though $S_2^\mu$ is not a conserved current anymore, we will assume that time translation invariance is not broken so that the time derivative vanishes $\partial_t\vev{S_2^0(t,\bb{x})}=0$. Since we allow space translation invariance to be broken, in principle there could be a non-zero current
\begin{equation}\label{magnetiz}
\vev{J^i(\bb{x})}=\epsilon^{ij}\partial_j M(\bb{x}),
\end{equation}
where $M(\bb{x})$ is the space-dependent part of the magnetization. This will enter in the spatial components of the angular momentum current
\begin{equation}
\partial_i \vev{S^i_2(\bb{x})} = -\partial_j\left(\frac{B}{2}\epsilon^{ij}x_i M(\bb{x})\right).
\end{equation}
Given that $S_2^i=\frac{1}{2}\epsilon_{jk}x^j T^{ik}$, by direct comparison we find that the expectation value of the stress tensor should be
\begin{equation}
\vev{T^{ij}(\bb{x})}=(P_0-B M(\bb{x}))\delta^{ij},
\end{equation}
where $P_0$ is a constant contribution to the pressure.

Because of the term depending on the current density in the conservation equation, we should also consider current-current and mixed retarded correlators
\begin{equation}
\begin{split}
G_R^{\mu\nu,\alpha}(t-\hat{t},\bb{x},\bb{\hat{x}})=i\Theta(t-\hat{t})\vev{\left[T^{\mu\nu}(t,\bb{x}),J^\alpha(\hat{t},\bb{\hat{x}})\right]},\\
G_R^{\alpha,\mu\nu}(t-\hat{t},\bb{x},\bb{\hat{x}})=i\Theta(t-\hat{t})\vev{\left[J^\alpha(t,\bb{x}),T^{\mu\nu}(\hat{t},\bb{\hat{x}})\right]},\\
G_R^{\mu\nu}(t-\hat{t},\bb{x},\bb{\hat{x}})=i\Theta(t-\hat{t})\vev{\left[J^{\mu}(t,\bb{x}),J^\nu(\hat{t},\bb{\hat{x}})\right]},
\end{split}
\end{equation}

Although due to the presence of the magnetic field the calculation is technically more involved, one can follow the same steps to derive the Ward identity as in the previous section. We have included the details in the Appendix~\ref{app:magn}. The original form \eqref{I13sol} is still valid, but  \eqref{I13solb} should be modified to
\begin{equation}\label{I13solB}
\boxed{
\begin{aligned}
\tilde{I}_{13}(\bb{\epsilon}) =&-i\omega\frac{\tilde{k}(2\bb{\epsilon})}{2}+ S_{ijkl}\frac{\partial}{\partial p_i}\frac{\partial}{\partial q_k}\left( \left[ p_n q_m \tilde{G}_R^{njml}(\omega,\bb{p},\bb{q})\right]_{\bb{p}=\bb{\epsilon},\bb{q}=-\bb{\epsilon}} \right.\\
&+i\omega B \left[  \epsilon^l_{\ m}\tilde{G}_R^{0j,m}(\omega,\bb{p},\bb{q})- \epsilon^j_{\ n}\tilde{G}_R^{n,0l}(\omega,\bb{p},\bb{q})\right]_{\bb{p}=\bb{\epsilon},\bb{q}=-\bb{\epsilon}}\\
&\left.-B^2\left[\delta^{jl}\delta_{nm}\tilde{G}_R^{nm}(\omega,\bb{p},\bb{q})-\tilde{G}_R^{lj}(\omega,\bb{p},\bb{q})\right]_{\bb{p}=\bb{\epsilon},\bb{q}=-\bb{\epsilon}} \right).
\end{aligned}}
\end{equation}
where $\tilde{k}$ substitutes the angular momentum density and is defined as
\begin{equation}
\tilde{k}(2\bb{\epsilon})=\tilde{\ell}(2\bb{\epsilon})-\frac{B}{2}\int d^2 \bb{x}\, e^{-i2\bb{\epsilon}\cdot\bb{x}}\, \bb{x^2}\vev{J^0(\bb{x})}.
\label{lB}
\end{equation}

\subsubsection{Adding dissipative terms}

So far we have assumed that momentum is conserved, except for the presence of an external magnetic field. This is not necessarily the case in many systems of interest, in particular if the microscopic theory is not translation invariant but there is a lattice or impurities with which the degrees of freedom that carry charge and momentum can scatter. A simple way to model the momentum loss is by adding drag terms to the conservation equation of momentum
\begin{equation}
\partial_\mu T^{\mu i}=B\epsilon^i_{\ j} J^j-\lambda_J J^i-\lambda_T T^{0i}.
\end{equation}
This is a purely phenomenological characterization of a system where the only effect of the scatterers is to change the momentum, as it happens in the Drude model. This kind of approximation has been used for instance in graphene \cite{Mendoza2013} and in the description of strongly coupled critical points  with a dilute concentration of impurities \cite{Hartnoll2007}. This simple approximation is not expected to hold in more general cases like impurities with magnetic momentum or when the scattering with the impurities is strong.

It is straightforward to repeat the same steps that we used to derive the Ward identity with the new terms. We find that \eqref{I13solb} should be modified to
\begin{equation}\label{I13solBdis}
\boxed{\begin{aligned}
\tilde{I}_{13}(\bb{\epsilon}) =& S_{ijkl}\frac{\partial}{\partial p_i}\frac{\partial}{\partial q_k} \left( \left[ p_n q_m \tilde{G}_R^{njml}(\omega,\bb{p},\bb{q})
-\lambda_T^2  \tilde{G}_R^{0j0l}(\omega,\bb{p},\bb{q})\right]_{\bb{p}=\bb{\epsilon},\bb{q}=-\bb{\epsilon}} \right.\\
&+(i\omega+\lambda_T)(B\epsilon^l_{\ m}-\lambda_J\delta^l_m)\left[\tilde{G}_R^{0j,m}(\omega,\bb{p},\bb{q})\right]_{\bb{p}=\bb{\epsilon},\bb{q}=-\bb{\epsilon}}\\
&+(-i\omega+\lambda_T)(B\epsilon^j_{\ n}-\lambda_J\delta^j_n)\left[\tilde{G}_R^{n,0l}(\omega,\bb{p},\bb{q})\right]_{\bb{p}=\bb{\epsilon},\bb{q}=-\bb{\epsilon}}\\
&\left.-(B\epsilon^j_{\ n}-\lambda_J\delta^j_n)(B\epsilon^l_{\ m}-\lambda_J\delta^l_m)\left[\tilde{G}_R^{nm}(\omega,\bb{p},\bb{q})\right]_{\bb{p}=\bb{\epsilon},\bb{q}=-\bb{\epsilon}} \right)\\
&+\text{contact terms}.
\end{aligned}}
\end{equation}
We do not write the contact terms explicitly because we will be interested mainly in using this formula in systems where expectation values are constant, where the contact terms will vanish.
\section{Relation between Hall viscosity and conductivities}
\label{sec:cond}

A non-zero angular momentum requires an expectation value of the momentum density of the form
\begin{equation}
\vev{T^{0i}(\bb{x})}=\frac{1}{2}\epsilon^{ij}\partial_j \ell(\bb{x}).
\end{equation}
This is compatible with rotational symmetry if $\ell$ is a function of $\bb{x^2}$.
The angular momentum is, after integrating by parts,
\begin{equation}
\vev{L_{xy}}=\int d^2\bb{x} \, \epsilon_{ij} x^i\vev{T^{0j}(\bb{x})}=\int d^2\bb{x} \ell(\bb{x^2}).
\end{equation}
Note however, that if translational invariance was exact, then necessarily $\vev{T^{0i}(\bb{x})}=0$ and $\vev{L_{xy}}=0$.

In  a translationally invariant theory correlators  have the form
\begin{equation}
\tilde{G}_R^{\mu\nu\alpha\beta}(\omega,\bb{p},\bb{q}) =(2\pi)^2\Gamma^{\mu\nu\alpha\beta}\left(\omega,\frac{\bb{p}+\bb{q}}{2}\right)\delta^{(2)}(\bb{p}-\bb{q}).
\end{equation}
We will define $\bb{P}=(\bb{p}+\bb{q})/2$ and $\bb{k}=\bb{p}-\bb{q}$.

Then \eqref{I13sol}  becomes
\begin{equation}
\begin{split}
\frac{\tilde{I}_{13}(\bb{\epsilon})}{(2\pi)^2} &=\frac{\omega^2}{4}\,S_{ijkl}\, \frac{\partial^2}{\partial P_i\partial P_k}\Gamma^{0j0l}(\omega,\bb{P})\delta^{(2)}(2\bb{\epsilon})\Bigg|_{\bb{P}=\bb{0}}\\
&+\frac{\omega^2}{2}\,S_{ijkl}\left[ \frac{\partial}{\partial P_k}\frac{\partial}{\partial k_i}-\frac{\partial}{\partial P_i}\frac{\partial}{\partial k_k}\right]\Gamma^{0j0l}(\omega,\bb{P})\delta^{(2)}(\bb{k})\Bigg|_{\bb{P}=\bb{0},\bb{k}=2\bb{\epsilon}}\\
&-\omega^2S_{ijkl}\Gamma^{0j0l}(\omega,\bb{0})\frac{\partial^2}{\partial k_i \partial k_k}\delta^{(2)}(\bb{k})\Bigg|_{\bb{k}=2\bb{\epsilon}}.
\end{split}
\end{equation}
Note that, if rotational invariance is not broken, the correlator can depend only on even powers of the momentum, $\bb{P}^2$ or $P^j P^l$. On the other hand, conservation of the energy-momentum tensor implies that $\omega^2\Gamma^{0j0l}(\omega,\bb{0})=0$. Therefore, the last two terms vanish and
\begin{equation}
\begin{split}
\frac{\tilde{I}_{13}(\bb{\epsilon})}{(2\pi)^2} &=\frac{\omega^2}{4}\,S_{ijkl}\, \frac{\partial^2}{\partial P_i\partial P_k}\Gamma^{0j0l}(\omega,\bb{P})\delta^{(2)}(2\bb{\epsilon})\Bigg|_{\bb{P}=\bb{0}}.
\end{split}
\end{equation}

\subsection{Systems with conserved momentum}

For a translationally invariant system \eqref{I13solb} becomes
\begin{equation}
\begin{split}
\frac{\tilde{I}_{13}(\bb{\epsilon})}{(2\pi)^2} &=\frac{1}{4}S_{ijkl}\frac{\partial^2}{\partial P_i \partial P_k}\left[P_n P_m \Gamma^{njml}(\omega,\bb{P})\right]_{\bb{P}=\bb{0}}\delta^{(2)}(2\bb{\epsilon}).
\end{split}
\end{equation}
We have used that, as a distribution,
\begin{equation}
\frac{\partial}{\partial k_i}\left[ k_n \cdots \delta^{(2)}(\bb{k})\right]\equiv 0,
\end{equation}
where the dots denote a polynomial on the components of $\bb{k}$.

Collecting all the terms proportional to $\delta^{(2)}(2\bb{\epsilon})$ we get
\begin{equation}\label{ward0}
(S_{ijkl}+S_{kjil}) \Gamma^{ijkl}(\omega,\bb{0})=\omega^2\,S_{ijkl}\, \frac{\partial^2}{\partial P_i\partial P_k}\Gamma^{0j0l}(\omega,\bb{P})\Bigg|_{\bb{P}=\bb{0}}.
\end{equation}
We present an alternative way to obtain the same result in the Appendix~\ref{app:alt}, where it is not necessary to use the regulator $\bb{\epsilon}$.

If there is rotational invariance then we can expand the stress-tensor correlator as
\begin{equation}\label{viscotens}
\begin{split}
\Gamma^{ijkl}(\omega,\bb{0}) &=-i\omega\left[\eta(\omega)(\delta^{ik}\delta^{jl}+\delta^{il}\delta^{jk}+( \zeta(\omega)-\eta(\omega))\delta^{ij}\delta^{kl}\right]\\
&-i\omega\frac{\eta_H(\omega)}{2}(\epsilon^{ik}\delta^{jl}+\epsilon^{il}\delta^{jk}+\epsilon^{jk}\delta^{il} +\epsilon^{jl}\delta^{ik}).
\end{split}
\end{equation}
Where the coefficients can be complex functions of the frequency. The real part of the coefficients are the usual transport coefficients. Note that viscosity terms describe the response of the system to time-dependent spatial deformations. In general, there is also a response to time-independent deformations (elastic response) determined by contact terms in the correlators, such as the inverse compressibility contribution in \cite{Bradlyn:2012ea}. Due to the contraction with the tensor $S_{ijkl}$, these terms will drop from our expressions, and we will neglect them in the following.

Introducing this in \eqref{ward0} we find the relation
\begin{equation}\label{etah1}
\eta_H(\omega)=-i\omega S_{ijkl}\frac{\partial^2}{\partial P_i\partial P_k}\Gamma^{0j0l}(\omega,\bb{P})\Bigg|_{\bb{P}=\bb{0}}.
\end{equation}

\subsection{Systems with a magnetic field}

If we use the expression \eqref{I13solB} for a constant magnetic field, we find
\begin{equation}\label{I13sol22Delta}
\begin{split}
\frac{\tilde{I}_{13}(\bb{\epsilon})}{(2\pi)^2} &=\frac{1}{4}(S_{ijkl}+S_{kjil}) \Gamma^{ijkl}(\omega,\bb{0})\delta^{(2)}(2\bb{\epsilon})\\
&+i \frac{\omega B}{4} S_{ijkl}\frac{\partial^2}{\partial P_i\partial P_k}\left[\epsilon^l_{\ m} \Gamma^{0j,m}(\omega,\bb{P})-\epsilon^j_{\ m}  \Gamma^{m,0l}(\omega,\bb{P}) \right]_{\bb{P}=\bb{0}}\delta^{(2)}(2\bb{\epsilon})\\
&-\frac{B^2}{4}S_{ijkl}\frac{\partial^2}{\partial P_i\partial P_k}\left[\delta^{jl}\delta_{nm}\Gamma^{nm}(\omega,\bb{P})-\Gamma^{lj}(\omega,\bb{P}) \right]_{\bb{P}=\bb{0}}\delta^{(2)}(2\bb{\epsilon}).
\end{split}
\end{equation}
The terms proportional to the second derivatives of the delta function cancel each other (see (\ref{SecDerB})).
We will now define the (complex) conductivities
\begin{equation}
\Gamma^{ij}=-i\omega \sigma^{ij}, \ \ \Gamma^{0i0j}=-i\omega \kappa^{ij}, \ \ \Gamma^{0i,j}=-i\omega \alpha^{ij}, \ \ \Gamma^{i,0j}=-i\omega \bar{\alpha}^{ij},
\end{equation}
where $\sigma$, $\kappa$ and $\alpha$, $\bar{\alpha}$ can be identified as the electric, ``momentum'' and ``mixed current-momentum'' conductivities respectively. In a relativistic system $T^{0i}=T^{i0}$ so $\kappa$ and $\alpha$ are combinations of thermal, thermoelectric and electric conductivities. There can also be a diamagnetic term in the correlator that describes the response to time-independent perturbations, but it will drop from our formulas after contracting with the tensor $S_{ijkl}$, so we will neglect it in the following.

The Hall electric and momentum conductivities are defined as
\begin{equation}
\sigma_H = \frac{1}{2}\epsilon_{ij}\sigma^{ij},\ \
\kappa_H = \frac{1}{2}\epsilon_{ij}\kappa^{ij}.
\end{equation}
We will also define the trace of the  mixed conductivity as $\tr\alpha=\delta_{ij}\alpha^{ij}$. Then, we can write the Ward identity for the Hall viscosity as\footnote{Details of the derivation are in Appendix~\ref{app:mag2}.}
\begin{equation}
\boxed{
\eta_H(\omega)=\omega^2\frac{\partial \kappa_H}{\partial \bb{P^2}}+B^2\frac{\partial \sigma_H}{\partial \bb{P^2}}+i\frac{\omega B}{2}\left[\frac{\partial \tr\alpha}{\partial \bb{P^2}} +\frac{\partial \tr\bar{\alpha}}{\partial \bb{P^2}}\right] \Bigg|_{\bb{P}=\bb{0}}.}
\label{etakappa}
\end{equation}
Or, in matrix form,
\begin{equation}
\eta_H(\omega)=\frac{\partial }{\partial \bb{P^2}}\left(
\begin{array}{cc}
\omega & B
\end{array}
\right)
\left(
\begin{array}{cc}
\kappa_H & \frac{i\tr\alpha}{2}\\
\frac{i\tr\bar{\alpha}}{2} & \sigma_H
\end{array}
\right)
\left(
\begin{array}{c}
\omega \\ B
\end{array}
\right)\Bigg|_{\bb{P}=\bb{0}}.
\end{equation}

\subsection{Adding dissipative terms}

The generalization of the formula above when momentum dissipation is included is straightforward using \eqref{I13solBdis} as starting point. The result is
\begin{equation}\label{etahfin}
\boxed{
\begin{aligned}
\eta_H(\omega)&=(\omega^2+\lambda_T^2)\frac{\partial \kappa_H}{\partial \bb{P^2}}+(B^2+\lambda_J^2)\frac{\partial \sigma_H}{\partial \bb{P^2}}\\
&+(i\omega+\lambda_T) \left[\frac{B}{2}\frac{\partial \tr\alpha}{\partial \bb{P^2}} +\lambda_J \frac{\partial\alpha_H}{\partial \bb{P^2}}\right]+(i\omega-\lambda_T) \left[\frac{B}{2}\frac{\partial \tr\bar{\alpha}}{\partial \bb{P^2}}-\lambda_J \frac{\partial\bar{\alpha}_H}{\partial \bb{P^2}}\right]\Bigg|_{\bb{P}=\bb{0}}.
\end{aligned}}
\end{equation}
We can also express the result in matrix form as
\begin{equation}
\begin{split}
&\eta_H(\omega)=\\
&\frac{\partial }{\partial \bb{P^2}}\left(
\begin{array}{cc}
\omega-i\lambda_T & B-i\lambda_J
\end{array}
\right)
\left(
\begin{array}{cc}
\kappa_H & i\frac{\frac{B}{2}\tr\alpha+\lambda_J \alpha_H }{B+i\lambda_J}\\
 i\frac{\frac{B}{2}\tr\bar{\alpha}-\lambda_J \bar{\alpha}_H }{B-i\lambda_J}& \sigma_H
\end{array}
\right)
\left(
\begin{array}{c}
\omega+i\lambda_T \\ B+i\lambda_J
\end{array}
\right)\Bigg|_{\bb{P}=\bb{0}}.
\end{split}
\end{equation}

\subsection{Galilean invariant theories}

Our results are completely general for relativistic and non-relativistic theories. Previous results \cite{Hoyos:2011ez, Bradlyn:2012ea, Hoyos:2013eha} were obtained in special cases with Galilean invariance and an universal charge to mass ratio. In such systems it is possible to make the following identification between the current and the momentum density \begin{equation}
T^{0i}=m J^i.
\end{equation}
For particles of unit charge and mass $m$.

Therefore, the $\kappa$ and $\alpha$ conductivities are not independent, but are related to the electric conductivity as
\begin{equation}
\kappa^{ij}=m^2 \sigma^{ij}, \ \ \alpha^{ij}=\bar{\alpha}^{ij}=m\sigma^{ij}.
\end{equation}
Starting with the most general expression \eqref{etahfin}, this leads to
\begin{equation}\label{etahnr}
\eta_H(\omega)=m^2\frac{\partial}{\partial \bb{P^2}}\left[(\omega^2+\omega_c^2+\lambda_{NR}^2)\sigma_H+i\omega\omega_c \tr\sigma\right]_{\bb{P}=\bb{0}}.
\end{equation}
Where $\omega_c=B/m$ is the cyclotron frequency and the non-relativistic drag coefficient is
\begin{equation}
\lambda_{NR}=\frac{\lambda_J}{m}+\lambda_T.
\end{equation}
Momentum dissipation was not considered previously, so in order to make a comparison we should set  $\lambda_{NR}=0$. Then, equation \eqref{etahnr} agrees with the Kubo formula (4.14) in \cite{Bradlyn:2012ea}.\footnote{Up to the sign of the imaginary part that is due to the conventions used to define the correlators.} This result, obtained from a formal derivation using the Ward identity, was checked in several examples in \cite{Bradlyn:2012ea} and is also confirmed by the effective field theory analysis of Hall systems \cite{Hoyos:2011ez} (where this type of relation was originally derived) and chiral superfluids \cite{Hoyos:2013eha}.

In a recent paper \cite{Geracie:2014} it was shown using non-relativistic diffeomorphism invariance that when parity is broken the relation between momentum and current can be modified to\footnote{For related works on the application of non-relativistic diffeomorphisms to effective theories see \cite{Gromov2014a} and also \cite{Gromov2014,Bradlyn2014} for systems without Galilean invariance.}
\begin{equation}
T^{0i}=m J^i-\frac{g-2s}{4}\epsilon^{ij}\partial_j J^0,
\end{equation}
where $g$ is the $g$-factor or gyromagnetic ratio that determines the coupling to an external magnetic field and $s$ determines the coupling to the spin connection. Using current conservation, this will modify the relation between conductivities to
\begin{align}
\kappa^{ij}&=m^2 \sigma^{ij}-im\frac{g-2s}{4\omega}P_k P_l \left[\epsilon^{jk}\sigma^{il}-\epsilon^{ik}\sigma^{lj} \right]+\frac{(g-2s)^2}{16 \omega^2}\epsilon^{ik}\epsilon^{jl}P_k P_l P_n P_m\sigma^{nm},\\
 \alpha^{ij} &=m\sigma^{ij}+i\frac{g-2s}{4\omega}\epsilon^{ik}P_k P_l\sigma^{lj},\\
 \bar{\alpha}^{ij} &=m\sigma^{ij}-i\frac{g-2s}{4\omega}\epsilon^{jk}P_k P_l\sigma^{il}.
\end{align}
Using these formulas in \eqref{etakappa}, it is straightforward to check that the Hall viscosity is shifted respect to \eqref{etahnr} by a term
\begin{equation}
\Delta\eta_H(\omega)=im\frac{g-2s}{4}\left[\frac{\omega}{2}\tr \sigma-i\omega_c \sigma_H \right],
\end{equation}
in agreement with the result of \cite{Geracie:2014}.

\section{Relation to angular momentum density}
\label{sec:ang}

In Quantum Hall systems and other topological states such as chiral superfluids, the Hall viscosity is proportional to the shift $\cS$ \cite{Tokatly2007,Read:2008rn,Tokatly2009,Read2011}. More precisely,
\begin{equation}\label{etahn}
\eta_H=\frac{\cS}{4}\bar{n},
\end{equation}
where $\bar{n}$ is the average particle number density. When put on a curved space, the shift determines the change in the number of particles relative to flat space
\begin{equation}
N=\nu^{-1}N_\phi-(1-g)\cS,
\end{equation}
where $g$ is the genus of the two-dimensional surface, $\nu$ is the filling fraction for a Hall system (for chiral superfluids $\nu^{-1}=0$) and $N_\phi$ is the number of magnetic flux quanta. In the superfluid $\cS$ is the orbital angular momentum of the Cooper pair. For free non-relativistic fermions in a magnetic field it is a mean orbital angular momentum per particle, defined as $\cS=2E_0/\omega_c$, where $E_0$ is the energy of the ground state and $\omega_c$ the cyclotron frequency \cite{Bradlyn:2012ea}.

In general the relation between Hall viscosity and angular momentum is not expected to hold, specially if the theory is gapless. An illustration of this are gauge/gravity models where the two quantities seem to be independent \cite{Liu:2014gto}. However, in \cite{Son:2013xra} it was found that for a relativistic $p$-wave superfluid the relation seems to be valid even at finite temperature.

Even though \eqref{etahn} may not hold in general, it is quite clear that in the presence of an angular momentum density there will be a contribution to the Hall viscosity. In the absence of magnetic field, from \eqref{I13solb} we have the relation
\begin{equation}
\begin{split}
0&=i\omega\frac{\tilde \ell(2\bb{\epsilon})}{2}+ S_{ijkl} \tilde{G}_R^{ijkl}(\omega,\bb{\epsilon},-\bb{\epsilon})-\omega^2S_{ijkl}\frac{\partial}{\partial p_i}\frac{\partial}{\partial q_k}\tilde{G}_R^{0j0l}(\omega,\bb{p},\bb{q})\Bigg|_{\bb{p}=\bb{\epsilon},\bb{q}=-\bb{\epsilon}}.
\end{split}
\end{equation}
We assume that rotational invariance is not broken but translation invariance can be. $\tilde\ell(\bb{0})$ equals the total angular momentum of the system, if it is finite. If the system is made of $N$ particles carrying an amount of angular momentum $\bar{\ell}$ on average, then $\tilde\ell(\bb{0})=N\bar{\ell}$. This diverges in the thermodynamic limit $N\to \infty$. In a system where the number of particles is not conserved and the density of angular momentum is approximately constant throughout space, then $\tilde\ell(\bb{0})$ has a volume divergence. In principle the same scaling with the volume is expected in the other terms.

Let us introduce the system in a finite volume $V_2$, in this case it is more convenient to work with the coordinate-dependent expressions
\begin{equation}
\begin{split}
0&=i\omega\frac{V_2\bar{\ell}}{2}+ S_{ijkl}\int_{V_2} d^2\bb{x}\int_{V_2}d^2\bb{\hat{x}} \,\tilde{G}_R^{ijkl}(\omega,\bb{x},\bb{\hat{x}})-\omega^2S_{ijkl}\int_{V_2} d^2\bb{x}\int_{V_2}d^2\bb{\hat{x}} x^i \hat{x}^k\,\tilde{G}_R^{0j0l}(\omega,\bb{x},\bb{\hat{x}}).
\end{split}
\end{equation}
Where the average angular momentum is defined as
\begin{equation}
\bar{\ell}=\frac{1}{V_2}\int_{V_2} d^2\bb{x} \ell(\bb{x}).
\end{equation}
The tensor structure of the correlator $\tilde{G}_R^{ijkl}(\omega,\bb{x},\bb{\hat{x}})=-i\omega \eta^{ijkl}(\omega,\bb{x},\bb{\hat{x}})$ is the same as in \eqref{viscotens}, but with coefficients that depend on the coordinates. We can define an average viscosity tensor as
\begin{equation}
\bar{\eta}^{ijkl}(\omega)=\frac{1}{V_2}\int_{V_2} d^2\bb{x}\int_{V_2}d^2\bb{\hat{x}} \,\eta^{ijkl}(\omega,\bb{x},\bb{\hat{x}}).
\end{equation}
If translation invariance was unbroken $\bar{\eta}$ would be the same as the zero momentum viscosity tensor.

Then, we find the following relation between the average Hall viscosity and the average angular momentum density
\begin{equation}
\boxed{
\bar{\eta}_H(\omega)=-\frac{\bar{\ell}}{2}-i\omega S_{ijkl}\frac{1}{V_2}\int_{V_2} d^2\bb{x}\int_{V_2}d^2\bb{\hat{x}} x^i \hat{x}^k\,\tilde{G}_R^{0j0l}(\omega,\bb{x},\bb{\hat{x}}).}
\label{etal}
\end{equation}

The confinement of the theory to finite volume could be due to an effective potential that depends on the scale $V_2=1/\delta$. The potential will break translation invariance, and should affect to the conservation equations of the energy-momentum tensor, but in the limit $\delta\to 0^+$ the breaking goes away and translation invariance is recovered. For simplicity we will assume that rotational invariance is not broken. In a situation like this, it should be possible to approximate the correlation function of the stress tensor as
\begin{equation}
\tilde{G}_R^{ijkl}(\omega,\bb{p},\bb{q})\simeq (2\pi)^2\Gamma^{ijkl}\left(\frac{\bb{p}+\bb{q}}{2} \right)\eta_\delta(\bb{p}-\bb{q})+\cdots,
\end{equation}
where $\eta_\delta(\bb{p}-\bb{q})$ is a function that becomes a Dirac delta when $\delta \to 0^+$. An example is
\begin{equation}
\eta_\delta(\bb{p}-\bb{q})=\frac{e^{-(\bb{p}-\bb{q})^2/(2\delta)}}{2\pi\delta}.
\end{equation}

If we now take the $\bb{\epsilon}\to \bb{0}$ limit,
\begin{equation}
\tilde{G}_R^{ijkl}(\omega,\bb{p},\bb{q})\simeq \frac{2\pi}{\delta}\Gamma^{ijkl}\left(\omega,\bb{0} \right)+\cdots,
\end{equation}
and similarly, $\tilde\ell(\bb{0})=(2\pi/\delta) \bar{\ell}+\cdots$. The dots are terms that will vanish in the limit where translation invariance is restored $\delta\to 0^+$. The function $\Gamma^{ijkl}$ can be expanded in the same way as in the translationally invariant case, so the term proportional to $1/\delta$ introduces the Hall viscosity.

Although in the end we are interested in the $\delta \to 0^+$ limit, we will show that in  a gapped system, for any finite $\delta$ the static Hall viscosity satisfies the relation with the angular momentum
\begin{equation}\label{staticHall}
\lim_{\omega\to 0} \eta_H(\omega)= -\frac{\bar{\ell}}{2}.
\end{equation}
Note that if the order of limits is changed, so  $\delta\to 0$ is taken {\em before}  $\bb{\epsilon}\to \bb{0}$, the angular momentum density should vanish and we should recover the formula that relates the Hall viscosity to the conductivities. In principle this does not mean there is any contradiction. Physically, taking the $\bb{\epsilon}\to \bb{0}$ limit in the correlators means probing the system at longer wavelengths, while the $\delta\to 0$ limit removes the scale at which translation invariance is broken to larger distances. We can expect then that if we study wavelengths much larger than the inverse of the gap or other scales but much smaller than $1/\sqrt{\delta}$, the relation between Hall viscosity and conductivities will be approximately valid, and this will improve as $\delta\to 0$ for larger wavelengths. On the other hand, if we consider wavelengths of the order of $1/\sqrt{\delta}$ or larger, we are sensitive to the breaking of translation invariance and the value of the static Hall viscosity is fixed. Moreover, the value of the Hall viscosity is independent of $\delta$ if the angular momentum density is kept constant, so the relation should be valid even when $\delta\to 0$.

\begin{figure}[ht!]
	\begin{center}
	\includegraphics[width=0.8\textwidth]{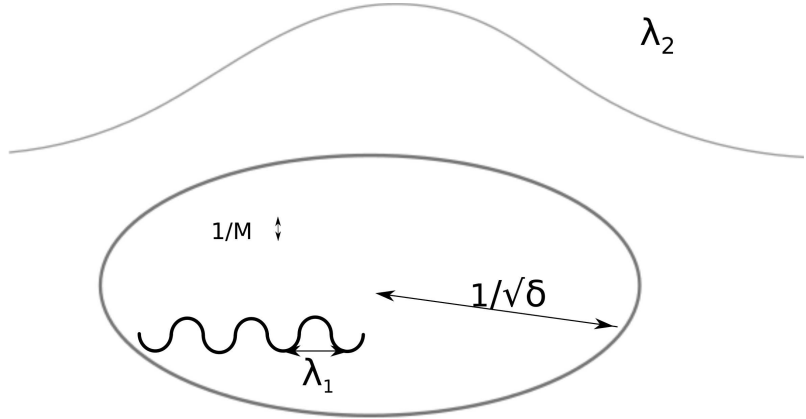}  \qquad\qquad
	\caption{Schematic picture of various length scales. If the wavelength satisfies $\frac{1}{M} \ll \lambda_1 \ll  \frac{1}{\sqrt{\delta}}$, where $M$ is the gap and $\sqrt{\delta}$ is the scale of translation symmetry breaking, the Hall viscosity can be related to the momentum derivative of Hall conductivity as \eqref{etakappa} or \eqref{etahnr}. If $\lambda_2 \gg  \frac{1}{\sqrt{\delta}}$, the Ward identity \eqref{etal} holds. On the one hand, taking the $\delta \to 0^+$ limit makes the relation to conductivity valid at zero momentum ($\lambda_1 \to \infty$). On the other hand, the value of the static Hall viscosity \eqref{staticHall} is independent of $\delta$. }
	\end{center}
	\end{figure}

\subsection{Systems with magnetic field}

In the case of free fermions in a magnetic field, the Hamiltonian has the form
\begin{equation}
\cH= \frac{\Pi_i^2}{2m}, \ \ \Pi_i=p_i-A_i.
\end{equation}
Where the `kinetic' momentum operators satisfy the commutation relations
\begin{equation}
[\Pi_i,\Pi_j]=i\epsilon_{ij}B, \ \ i[\cH,\Pi_i]=\omega_c\epsilon_i^{\ j}\Pi_j.
\end{equation}
In the rotationally invariant gauge the single-particle wavefunctions in the lowest Landau level can be expanded in a basis ($z=x+iy$)
\begin{equation}
\psi_n(z)=N_n z^n e^{-B|z|^2/4}.
\end{equation}
For the usual definition of the angular momentum operator $L_{xy}^p=xp_y-y p_x$ with $p_i$ the canonical momentum operators, these wavefunctions carry $n$ units of angular momentum. The total momentum of $N$ fermions in the lowest Landau level will be then of order $N^2$. However, $L_{xy}^p$ is not gauge-invariant and has no direct physical interpretation. A gauge-invariant definition involves the kinetic momentum operator $L_{xy}^\Pi=x\Pi_y-y\Pi_x$. For this operator, the angular momentum is independent of $n$ and is actually $-1$. For the $\cN$th Landau level, the single particle states have angular momentum $-(2\cN+1)$. In the case of $\nu$ filled Landau levels we can use the fact that each Landau level is equally degenerate, so the average value is
\begin{equation}
\frac{L_{xy}^\Pi}{N} =-\frac{1}{\nu}\sum_{\cN=0}^{\nu-1} (2\cN+1)=-\nu.
\end{equation}
These are the values that determine the shift.

In our analysis $T^{0i}$ is a gauge-invariant operator, we can see that it is indeed related to the kinetic momentum operators in quantum mechanics. In the presence of the magnetic field the conservation equation is $\partial_\mu T^{\mu i}=B\epsilon^i_{\ j}J^j$.
In addition, in a theory with Galilean invariance $T^{0i}=m J^i$, in which case we can write the conservation equation as
\begin{equation}
\partial_\mu T^{\mu i}=\omega_c\epsilon^i_{\ j} T^{0j},
\end{equation}
The momentum operators $P^i =\int d^2\bb{x} T^{0i}$ then satisfy
\begin{equation}
\partial_t P^i = \omega_c \epsilon^i_{\ j} P^j \ \ \Rightarrow \ \ i[\cH,P^i]=\omega_c \epsilon^i_{\ j} P^j,
\end{equation}
where $\cH$ is the Hamiltonian. This agrees with the commutation relation for the kinetic momentum operators $\Pi_i$. Therefore, the angular momentum $L_{xy}=\int d^2\bb{x} \epsilon_{ij}x^i T^{0j}$ corresponds to $L_{xy}^\Pi$ and should capture the right value of the shift.

From \eqref{lB} we see that in the presence of a magnetic field not only the angular momentum contributes but there is a term which would be divergent in the infinite volume limit if the density remains constant. This divergence is related to the static Hall conductivity. In the presence of the magnetic field we may extract a contribution from the current correlator of the form
\begin{equation}
G_R^{ij} (\omega;\bb{x},\bb{\hat{x}})=i\omega \epsilon^{ij}\frac{\bar{n}}{B} \delta^{(2)}(\bb{x}-\bb{\hat{x}})+\hat{G}_R^{ij} (\omega;\bb{x},\bb{\hat{x}}),
\end{equation}
where $\bar{n}$ is the average charge density. This leads to
\begin{equation}
\begin{split}
\tilde{I}_{13}(\bb{\epsilon}) &=
\int d^2 \bb{x}\, e^{-i2\bb{\epsilon}\cdot\bb{x}}\,i\omega\left[\vev{S_2^0(\bb{x})}-\frac{B}{4}x^2\left(\vev{J^0(\bb{x})}-\bar{n}\right)\right]\\
&+\int d^2 \bb{x}\,\, d^2 \bb{\hat{x}} \, e^{-i\bb{\epsilon}\cdot\bb{x}-i\bb{\epsilon}\cdot\bb{\hat{x}}} \,S_{ijkl}x^i\hat{x}^k\left( \partial_n\partial_{\hat{m}} G_R^{njml}(\omega;\bb{x},\bb{\hat{x}})-i\omega B\epsilon^j_{\ n} G_R^{n,0l}(\omega;\bb{x},\bb{\hat{x}})\right.\\
&\left.+i\omega B\epsilon^l_{\ m} G_R^{0j,m}(\omega;\bb{x},\bb{\hat{x}})
-B^2\epsilon^j_{\ n}\epsilon^l_{\ m} \hat{G}_R^{nm} (\omega;\bb{x},\bb{\hat{x}})\right).
\end{split}
\end{equation}
So the contact term vanishes when the density is constant $\vev{J^0(\bb{x})}=\bar{n}$.
There can also be a diamagnetic term in the current correlator $G_R^{ij} \sim \bar{n}\delta^{ij}$, but it will drop after contracting with $S_{ijkl}$.

\subsection{Spectral representation}

In \cite{Bradlyn:2012ea} it was argued that the static Hall viscosity will be exactly \eqref{etahn} in a system with a mass gap and no magnetic field. The argument uses the spectral representation of correlators, we generalize it to other field theories. First, let us define the vector operators
\begin{equation}
V^i_1=T^{0i}, \ \ V^i_2=\epsilon^i_{\ n}J^n.
\end{equation}
Then, the Ward identity for the average Hall viscosity can be written as
\begin{equation}\label{etahk}
\begin{split}
\bar{\eta}_H(\omega)&=-\frac{\bar{k}}{2}
+S_{ijkl}\frac{1}{V_2}\int_{V_2} d^2\bb{x}\int_{V_2} d^2\bb{\hat{x}} x^i \hat{x}^k
M^{ab} \vev{V_a^j  V_b^l }_R,
\end{split}
\end{equation}
where $\bar{k}$ is the full contact term  ($\bar{k}=\bar{\ell}$ if $B=0$) and
\begin{equation}
M^{ab}=\left(
\begin{array}{cc}
-i\omega & - B\\
 B & -i\frac{B^2}{\omega}
\end{array}
\right).
\end{equation}
Using the usual relations between correlators (see Appendix~\ref{app:spect}) we find ($\epsilon \to 0^+ $)
\begin{equation}
\begin{split}
& \vev{V_a^j  V_b^l }_R
 =2i \int \frac{d k_0}{2\pi}\frac{1}{\omega-k_0-i\epsilon}\rho_{ab}^{jl}(k_0,\bb{x},\bb{\hat{x}}).
 \end{split}
\end{equation}
Where $\rho_{ab}^{ij}$ is the spectral function.

There are no divergences as $\omega\to 0$ coming from the integral over $k_0$ as long as
\begin{equation}
\lim_{\omega\to 0} \rho^{ij}_{ab}(\omega,\bb{x},\bb{\hat{x}})<\infty.
\end{equation}
This can be checked from the decomposition of the pole in the principal value part and a delta function
\begin{equation}
\frac{1}{k_0-\omega+i\epsilon}= \cP \frac{1}{k_0-\omega}-i\pi \delta(k_0-\omega).
\end{equation}
Then, we find that
\begin{equation}
\int\frac{d k_0}{2\pi}\frac{1}{\omega-k_0-i\epsilon}\,\rho_{ab}^{jl}(k_0,\bb{x},\bb{\hat{x}})=\frac{i}{2}\rho_{ab}^{jl}(\omega ,\bb{x},\bb{\hat{x}})  -\cP\int\frac{d k_0}{2\pi}\frac{1}{k_0-\omega}\,\rho_{ab}^{jl}(k_0,\bb{x},\bb{\hat{x}}).
\end{equation}
Both the principal value and the imaginary term contribute to the real part of the Hall viscosity. Let us define
\begin{equation}
F_{ab}(k_0)=\frac{1}{V_2}S_{ijkl}\int_{V_2} d^2\bb{x}\int_{V_2} d^2\bb{\hat{x}}\,x^i\hat{x}^k\,\rho_{ab}^{jl}(k_0,\bb{x},\bb{\hat{x}}).
\end{equation}
Then, for the real part we have
\begin{equation}
{\rm Re}\,\bar{\eta}_H(\omega)=-\frac{\bar{k}}{2}-2\omega \cP\int\frac{d k_0}{2\pi}\frac{F_{11}(k_0)}{k_0-\omega}-\frac{2B}{\omega}\cP\int\frac{d k_0}{2\pi}\frac{F_{22}(k_0)}{k_0-\omega}+B\left(F_{12}(\omega)-F_{21}(\omega) \right).
\end{equation}
The expression for the imaginary part is
\begin{equation}
{\rm Im}\,\bar{\eta}_H(\omega)=i\omega F_{11}(\omega)+\frac{iB}{\omega}F_{22}(\omega)+2iB\cP\int\frac{d k_0}{2\pi}\frac{F_{12}(k_0)-F_{21}(k_0) }{k_0-\omega}.
\end{equation}
In the absence of a magnetic field, the formula for the Hall viscosity takes the simpler form
\begin{equation}
\boxed{
\bar{\eta}_H(\omega)=-\frac{\bar{\ell}}{2}-2\omega \cP\int\frac{d k_0}{2\pi}\frac{F_{11}(k_0)}{k_0-\omega}+i\omega F_{11}(\omega).}
\label{simpler}
\end{equation}
If $\omega F_{11}(\omega)\to 0 $ as $\omega\to 0$, then only the angular momentum density will contribute to the Hall viscosity. This will happen if there is an energy gap in the spectrum. Indeed  the spectral function can be formally expanded as a sum over energy eigenstates of the form
\begin{equation}
\rho_{11}^{ij}(\omega,\bb{x},\bb{\hat{x}})=2\pi\sum_{\alpha\neq 0}\delta(\omega-\varepsilon_\alpha) \,{\rm Im}\,\left(\bra{0}T^{0i}(\bb{x})\ket{\alpha}\bra{\alpha} T^{0j}(\bb{\hat{x}})\ket{0}\right),
\end{equation}
where $\ket{\alpha}$ are the energy eigenstates and $\varepsilon_\alpha$ is the energy difference with the ground state $\ket{0}$. Clearly, for $\varepsilon_\alpha\neq 0$ the function $\rho_{11}^{ij}(\omega,\bb{x},\bb{\hat{x}})$ vanishes at $\omega=0$. Note that there are no special requirements on the form of the spectrum above the gap. The situation is different at finite temperature, where the spectral function has the form (see Appendix~\ref{finiteT})
\begin{equation}
\begin{split}
&\rho_{ab}^{ij}(\omega,\bb{x},\bb{\hat{x}})_T\\
&=\pi\sum_{\alpha,\beta; \varepsilon_\alpha\neq \varepsilon_\beta}\left[ e^{-\varepsilon_\beta/T}\delta(\omega-(\varepsilon_\alpha-\varepsilon_\beta)) \,{\rm Im}\,\left(\bra{\beta} V_a^i(\bb{x})\ket{\alpha}\bra{\alpha}V_b^j(\bb{\hat{x}})\ket{\beta}\right)\right.\\
&\left.+ e^{-\varepsilon_\alpha/T}\delta(\omega+(\varepsilon_\alpha-\varepsilon_\beta)) \,{\rm Im}\,\left(\bra{\beta}V_b^j(\bb{\hat{x}})\ket{\alpha}\bra{\alpha} V_a^i(\bb{x})\ket{\beta}\right)\right].
\end{split}
\end{equation}
We see that if the spectrum is discrete the spectral function will vanish at zero frequency even at non-zero temperature. However, for a continuous spectrum this does not need to be true in general.

In conclusion, in the absence of magnetic fields, for any field theory with an energy gap, the static Hall viscosity at zero temperature will be given by Read's formula
\begin{equation}
\eta_H=-\frac{\bar{\ell}}{2}.
\end{equation}
If the theory does not have a gap, the relation depends on the matrix elements of $T^{0i}$. In a theory with spontaneous breaking we can have massless Goldstone bosons separated by an energy gap from other kind of excitations. In such a case the energy-momentum tensor at low energies will be proportional to derivatives of the Goldstone field $\phi$
\begin{equation}
T^{0i}\simeq \partial^0\phi \partial^i\phi,
\end{equation}
in which case one expects the matrix element of the momentum density to be proportional to the energy of the eigenstates
\begin{equation}
\bra{0}T^{0i}\ket{\alpha} \simeq i\varepsilon_\alpha \bra{0}\phi\partial^i\phi\ket{\alpha}.
\end{equation}
Even though the continuous of excitations of the Goldstone bosons reaches zero energy, this factor would prevent them from contributing to the Hall viscosity.
\section{Conformal transformations on the plane}
\label{sec:conf}

We have seen how the angular momentum density enters as a contact term in the Ward identity for area-preserving transformations, as was also derived by \cite{Bradlyn:2012ea} for non-relativistic theories. In general we expect it to enter in any relation between symmetry generators that includes rotations. As we will show now, this is indeed the case of spatial conformal transformations.

The commutator of a spatial translation with a special spatial conformal transformation is proportional to a spatial dilatation plus a spatial rotation
\begin{equation}
[P_i,K_j]=-2i M_{ij}+2i \delta_{ij} D_s.
\end{equation}
A representation of the algebra is
\begin{equation}
P_i=-i\partial_i, \ \ K_i=-i(\bb{x}^2\partial_i-2x_ix^k\partial_k), \ \ D_s=-ix^k\partial_k, \ \ M_{ij}=-i(x_i\partial_j-x_j\partial_i).
\end{equation}
If the expectation value of the angular momentum $\vev{M_{ij}}=\vev{L_{xy}}\epsilon_{ij}$ is non-zero, the commutator of $K_i$ and $P_i$ will be non-zero as well
\begin{equation}
\frac{1}{2}\epsilon^{ij}\vev{[P_i,K_j]}=-2i\vev{L_{xy}}.
\end{equation}
This implies the following Ward identity for the equal time commutator
\begin{equation}\label{ward3}
\frac{1}{2}\epsilon_{ij}\vev{\left[T^{0i}(t,\bb{x}), \kappa^{0j}(t,\bb{\hat{x}})\right]}= -4i\vev{S_2^0(t,\bb{x})}\delta^{(2)}(\bb{x}-\bb{\hat{x}}).
\end{equation}
Where we have defined the current density
\begin{equation}
\kappa^{\mu i}(t,\bb{x})=\bb{x}^2 T^{\mu i}(t,\bb{x})-2x^i x_kT^{\mu k}(t,\bb{x}).
\end{equation}
Note that
\begin{equation}
\partial_\mu \kappa^{\mu i} = -2 x^i T^k_{\ k},
\end{equation}
is not conserved.

Analogously to the derivation for area-preserving transformations, we will consider time derivatives of the retarded correlator
\begin{equation}
G_{PK}(t-\hat{t},\bb{x},\bb{\hat{x}})=i\Theta(t-\hat{t})\frac{1}{2}\epsilon_{ij}\vev{\left[T^{0i}(t,\bb{x}), \kappa^{0j}(\hat{t},\bb{\hat{x}})\right]}.
\end{equation}
For this, we will define
\begin{equation}
I_{PK}(\bb{\epsilon})=\partial_{\hat{t}}\int d^2\bb{x}d^2\bb{\hat{x}}e^{-i\bb{\epsilon}\cdot\bb{x}-i\bb{\epsilon}\cdot\bb{\hat{x}}}  G_{PK}(t-\hat{t},\bb{x},\bb{\hat{x}}).
\end{equation}
More explicitly,
\begin{equation}
I_{PK}(\bb{\epsilon}) =\frac{1}{2}\epsilon_{ij}\partial_{\hat{t}}\int d^2\bb{x}d^2\bb{\hat{x}}e^{-i\bb{\epsilon}\cdot\bb{x}-i\bb{\epsilon}\cdot\bb{\hat{x}}}\left[\bb{\hat{x}^2}\delta^j_k-2\hat{x}^j \hat{x}_k\right] G_R^{0i0k}(t-\hat{t},\bb{x},\bb{\hat{x}}) .
\end{equation}
We now Fourier transform $I_{PK}$ with respect to $t-\hat{t}$ and use the form of the correlators in momentum space
\begin{equation}
\tilde{I}_{PK}(\bb{\epsilon}) =\frac{i\omega}{2}\epsilon_{ij}\left[\delta^j_k\frac{\partial^2}{\partial q_l \partial q^l}-2\frac{\partial^2}{\partial q_j \partial q^k}\right] G_R^{0i0k}(\omega,\bb{\epsilon},\bb{q})\Bigg|_{\bb{q}=-\bb{\epsilon}} .
\end{equation}
If we use the explicit form of the retarded correlator, the conservation of the energy-momentum tensor and the Ward identity \eqref{ward3}, we find the following expression
\begin{equation}
\begin{split}
I_{PK}(\bb{\epsilon}) &=-4\delta(t-\hat{t})\int d^2\bb{x}e^{-2i\bb{\epsilon}\cdot\bb{x}} \vev{S_2^0(\bb{x})}\\
&-\frac{1}{2}\epsilon_{ij}\int d^2\bb{x}d^2\bb{\hat{x}}e^{-i\bb{\epsilon}\cdot\bb{x}-i\bb{\epsilon}\cdot\bb{\hat{x}}}\left[\bb{\hat{x}^2}\delta^j_k-2\hat{x}^j \hat{x}_k\right] \partial_{\hat{m}}G_R^{0imk}(t-\hat{t},\bb{x},\bb{\hat{x}}).
\end{split}
\end{equation}
We now Fourier transform
\begin{equation}
\begin{split}
\tilde{I}_{PK}(\bb{\epsilon}) &=-2\tilde{\ell}(2\bb{\epsilon})
+\frac{i}{2}\epsilon_{ij}\left[\delta^j_k\frac{\partial^2}{\partial q_l \partial q^l}-2\frac{\partial^2}{\partial q_j \partial q^k}\right]  q_m \tilde{G}_R^{0imk}(\omega,\bb{p},\bb{q})\Bigg|_{\bb{p}=\bb{\epsilon},\bb{q}=-\bb{\epsilon}}.
\end{split}
\end{equation}
This leads to the relation
\begin{equation}\label{conformal}
\boxed{
\begin{aligned}
2 \tilde{\ell}(2\bb{\epsilon})=\frac{i}{2}\epsilon_{ij}\left[\delta^j_k\frac{\partial^2}{\partial q_l \partial q^l}-2\frac{\partial^2}{\partial q_j \partial q^k}\right]\left(  q_m\tilde{G}_R^{0imk}(\omega,\bb{p},\bb{q}) -\omega\tilde{G}_R^{0i0k}(\omega,\bb{p},\bb{q})\right)_{\bb{p}=\bb{\epsilon},\bb{q}=-\bb{\epsilon}}.
\end{aligned}}
\end{equation}

\subsection{Conformal Ward identity}

In addition to the Hall viscosity, we can also discuss the consequences of the Ward identity derived from spatial conformal transformations \eqref{conformal}.
To leading order in $\bb{\epsilon}$, we have non-zero contributions coming from terms of the form
\begin{equation}
\begin{split}
\tilde{G}_R^{0imk}(\omega,\bb{p},\bb{q}) =& i\delta^{mk}\epsilon^{il}q_l \chi_\Omega+\cdots,\\
\tilde{G}_R^{0i0k}(\omega,\bb{p},\bb{q}) =& i\left(\epsilon^{il}q_l q^k+ \epsilon^{kl}q_l q^i\right)\Pi_\Omega+\cdots.
\end{split}
\end{equation}
Where the dots denote other terms with a different tensor structure that do not contribute to the Ward identity. The Ward identity becomes
\begin{equation}
-2 \tilde{\ell}(2\bb{\epsilon})+4\left(\chi_\Omega-\omega\Pi_\Omega \right)=0 .
\end{equation}
Then, normalizing by the volume, the coefficient $\chi_\Omega$ is
\begin{equation}
\chi_\Omega = \frac{\bar{\ell}}{2}+O(\omega).
\end{equation}
Note that if translation invariance was unbroken $\bar{\ell}=0$ and $\chi_\Omega=0$ at zero frequency. The value of $\chi_\Omega$ is related to the Hall bulk viscosity term in parity-breaking fluids. This term is relevant in fluids where the vorticity of the fluid $\Omega$ is non-zero. For a relativistic fluid with three-velocity $u^\mu$, $\Omega=-\epsilon^{\mu\nu\lambda}u_\mu\partial_\nu u_\lambda$. For small velocities $\Omega=\epsilon^{ij}\partial_i v_j$. The Hall bulk viscosity term at small velocities takes the form
\footnote{This is in the `magneticovortical' frame, where the value of the energy in the perfect fluid energy-momentum tensor is shifted by the magnetic field and the vorticity, see \cite{Jensen:2011xb} for details.}
\begin{equation}
T^{ij}_{H\,\text{bulk}}=-\tilde{x}_\Omega\delta^{ij}\Omega
\end{equation}
Then, to leading order in derivatives, the change in the stress tensor due to the vorticity is
\begin{equation}
\delta T^{ij} = \left( \frac{\partial P}{\partial \Omega}-\tilde{x}_\Omega\right)\delta^{ij}\Omega,
\end{equation}
where $P$ is the pressure appearing in the energy-momentum tensor at the ideal order. As was discussed in \cite{Jensen:2011xb}, in a translationally invariant system
\begin{equation}
 \frac{\partial P}{\partial \Omega}-\tilde{x}_\Omega=0.
\end{equation}
Therefore, the total change in the pressure, understood as the trace of the stress tensor, is actually zero. In view of this, if we define the total pressure to be $p_T=\delta_{ij}\vev{T^{ij}}/2$, the conformal Ward identity that we have derived implies that
\begin{equation}
\boxed{
\frac{\partial p_T}{\partial \Omega}=\frac{\bar{\ell}}{2}.}
\label{pressurerelation}
\end{equation}

In \cite{Jensen:2011xb} it was argued that $\frac{\partial P}{\partial \Omega}$ is also related to angular momentum density.\footnote{We thank Kristan Jensen for discussions about this point.} From the analysis of the hydrostatic generating functional \cite{Jensen:2012jh,Banerjee:2012iz} one can see that in the presence of space-dependent metric and gauge fields, there is a contribution that generates angular momentum. In the generating functional it appears as a term of the form
\begin{equation}
W_{\Omega}[g_{\mu\nu},A_\mu]=\int d^2x\, c_1\left(g_{00},A_0\right)\epsilon^{ij}\partial_i g_{0j}.
\end{equation}
Where the vorticity is $\Omega =\epsilon^{ij}\partial_i g_{0j}$.
The variation with respect to $g_{0i}$ leads to
\begin{equation}
T^{0i}_\Omega = \epsilon^{ij}\partial_j c_1,
\end{equation}
and $c_1$ can be identified with $\frac{\partial P}{\partial \Omega}$. Therefore $\frac{\partial P}{\partial \Omega}$ is proportional to the angular momentum density. Note that nevertheless it is necessary to break translation invariance in order to have a non-zero angular momentum, understood as the expectation value of the angular momentum operator. In principle one could give an alternative definition of `angular momentum density' in terms of the contact terms that appear in two-point functions.

In the derivation of the generating functional it is assumed that in the absence of sources the ground state will be translationally invariant. This makes a difference with the analysis we have made where the angular momentum is generated spontaneously. Note that $W_\Omega$ is independent of $g_{ij}$, so there is no contribution to the stress tensor from this term $T^{ij}_\Omega=0$, while we found that the stress tensor depends on the vorticity \eqref{pressurerelation} if the ground state is not translationally invariant in the absence of sources.

\section{Discussion}
\label{sec:disc}
In this work we derived a Ward identity relation between Hall viscosity and Hall conductivities that is valid for general relativistic
or non-relativistic $2+1$ dimensional quantum field theories.
The relation reduces to the known results of Galilean invariant theories \cite{Bradlyn:2012ea}, including the very recent result of \cite{Geracie:2014}.
We further generalized the relation by adding the effect of a drag viscous term.
 It would be of interest to verify these identities in explicit models. One suitable setup is the holographic $p$-wave superfluid \cite{Gubser:2008zu} for which the Hall viscosity was calculated in \cite{Son:2013xra}.

Ward identities introduce constraints among transport coefficients, which should be valid
in holographic models that have quantum field theory dual descriptions. Since most holographic models in the context of the
gauge/gravity correspondence  have a bottom-up description, the Ward identities
can be used to test which of these models may have a field theory dual. A more interesting relation is between Hall viscosity and conductivities, since the relation to angular momentum can be shown to hold only in very special cases.

The analysis of Ward identities that we carried out
can be extended to other linear transformations. In  Galilean invariant theories they give relations between shear and bulk viscosities and other components of the conductivity. We expect that a similar generalizations will apply for relativistic as well as non-boost invariant theories, such as Lifshitz field theories \cite{Ardonne:2003,Kachru:2008yh,Hoyos:2013eza}.

We showed that the relation between Hall viscosity and angular momentum density holds in special cases, i.e. at zero temperature and for gapped systems. In order to show the relation it was necessary to break translation invariance and take a limit.
This deserves additional analysis as there may be some subtleties, in particular if the energy of some states approaches the energy of the ground state in this limit.

We argued that the relation can be expected to hold also for systems with spontaneous breaking of symmetry, if there is a gap between the Goldstone bosons and other excitations. This is in agreement with results from effective field theory.  It would be interesting to have a more rigorous proof of this including an
explicit verification in particular models.

For systems with a background magnetic field, the relation between Hall viscosity and angular momentum is modified. This
is expected since in the static limit the conductivity terms entering the Ward identity should give some contributions as well. Such a modification
has already been discussed in the Galilean invariant cases \cite{Bradlyn:2012ea}. Since the Hall viscosity is related to the shift, which is topologically protected, it would be interesting to identify all the terms that enter in the Hall viscosity. They may be related to the existence of gapless modes.

We introduced nonzero temperature in the analysis.
In comparison to the zero temperature case, here not only the energy difference with respect to the ground state is relevant, but also the energy differences among all excited states. While the matrix elements are suppressed by factors of the frequency as at zero temperature, there are potential contributions
 to the Hall viscosity at non-zero magnetic field or if the density of states grows at small frequencies.
It would also be interesting to further study these cases.

\section*{Acknowledgements}
We would like to thank Moshe Goldstein for valuable discussions and Kristan Jensen and Nicholas Read for useful comments.
This work is supported in part by the I-CORE program of Planning and Budgeting Committee (grant number 1937/12), and by the US-Israel Binational Science Foundation (BSF).

\appendix

\section{Details of the calculation of Ward identities}
\label{sec:app1}

In this appendix we collect technical results and useful formulae that we used in the derivation of the Ward identities in section \ref{sec:ward}, and give an alternative derivation in the absence of magnetic fields and dissipation in \S~\ref{app:alt}. In all cases we used the following algebraic relations for the tensor $S$:
\begin{equation}\label{Srels}
\begin{split}
S_{ijkl}=S_{jikl}, \ \ S_{ijkl}=S_{ijlk}, \ \ \delta^{ij}S_{ijkl}=0, \ \ \delta^{kl} S_{ijkl}=0,\\
S_{ijkl}\delta^{ik}=-\frac{1}{4}\epsilon_{jl},\ \ S_{ijkl}\delta^{jl}=-\frac{1}{4}\epsilon_{ik}, \ \ S_{ijkl}\epsilon^{ik}=-\frac{1}{4}\delta_{jl},\ \ S_{ijkl}\epsilon^{jl}=-\frac{1}{4}\delta_{ik}.
\end{split}
\end{equation}

\subsection{Systems with conserved momentum}
\label{app:cons}

Written explicitly, the Fourier transform \eqref{FI13} is
\begin{equation}
\begin{split}
&\tilde{I}_{13}(\bb{\epsilon})=\omega^2\,S_{ijkl}\int d^2 \bb{x}\,\, d^2 \bb{\hat{x}} \frac{ d^2 \bb{p} d^2 \bb{q}}{(2\pi)^4} \,e^{-i\bb{\epsilon}\cdot\bb{x}-i\bb{\epsilon}\cdot\bb{\hat{x}}} \, \frac{\partial}{\partial p_i}\frac{\partial}{\partial q_k}\left[e^{i\bb{p}\cdot\bb{x}-i\bb{q}\cdot\bb{\hat{x}}}\right]\tilde{G}_R^{0j0l}(\omega,\bb{p},\bb{q}).
\end{split}
\end{equation}
Where we have used that
\begin{equation}
x^i \hat{x}^k e^{i\bb{p}\cdot\bb{x}-i\bb{q}\cdot\bb{\hat{x}}}=\frac{\partial}{\partial p_i}\frac{\partial}{\partial q_k}e^{i\bb{p}\cdot\bb{x}-i\bb{q}\cdot\bb{\hat{x}}}.
\end{equation}
We now integrate by parts the derivatives with respect to $p_q$ and $q_k$ and perform the integrals over space.\footnote{One can show that the boundary terms vanish by doing the integrals over space.} This gives a factor $\delta^{(2)}(\bb{p}-\bb{\epsilon})\delta^{(2)}(\bb{q}+\bb{\epsilon})$ that we use to compute the integrals over momentum. The result is \eqref{I13sol}.

From \eqref{I13b} we get
\begin{equation}
\begin{split}
\tilde{I}_{13}(\bb{\epsilon}) &=i\omega\int d^2 \bb{x}\,e^{-2i\bb{\epsilon}\cdot\bb{x}} \, \vev{S^0_2(\bb{x})}\\
&+S_{ijkl}\int d^2 \bb{x}\,\, d^2 \bb{\hat{x}} \frac{ d^2 \bb{p} d^2 \bb{q}}{(2\pi)^4} \,e^{-i\bb{\epsilon}\cdot\bb{x}-i\bb{\epsilon}\cdot\bb{\hat{x}}} \, p_n q_m\frac{\partial}{\partial p_i}\frac{\partial}{\partial q_k}\left[e^{i\bb{p}\cdot\bb{x}-i\bb{q}\cdot\bb{\hat{x}}}\right]\tilde{G}_R^{njml}(\omega,\bb{p},\bb{q}).
\end{split}
\end{equation}
Integrating by parts the derivatives with respect to momentum and doing first the integrals over the space directions and then over momentum we get \eqref{I13solb}.

\subsection{Systems with a magnetic field}
\label{app:magn}

As in the previous case, first we compute two time derivatives of the retarded correlator $G_{13}$ (defined with $S_a^\mu$) integrated over space
\begin{equation}
\begin{split}
I_{13}(\bb{\epsilon}) &=
\int d^2 \bb{x}\,\, d^2 \bb{\hat{x}} \, e^{-i\bb{\epsilon}\cdot\bb{x}-i\bb{\epsilon}\cdot\bb{\hat{x}}}S_{ijkl}x^i\hat{x}^k\left[-i\delta'(t-\hat{t})\vev{[T^{0j}(t,\bb{x}),T^{0l}(t,\bb{\hat{x}})]}\right.\\
&-2i\delta(t-\hat{t})\partial_t \vev{[T^{0j}(t,\bb{x}),T^{0l}(t,\bb{\hat{x}})]}\\
&+ \partial_n\partial_{\hat{m}} G_R^{njml}(t-\hat{t};\bb{x},\bb{\hat{x}})-B\epsilon^j_{\ n} \partial_{\hat{m}}G_R^{n,ml}(t-\hat{t};\bb{x},\bb{\hat{x}})\\
&\left.-B\epsilon^l_{\ m} \partial_{n}G_R^{nj,m}(t-\hat{t};\bb{x},\bb{\hat{x}})
+B^2\epsilon^j_{\ n}\epsilon^l_{\ m} G_R^{nm} (t-\hat{t};\bb{x},\bb{\hat{x}}) \right].
\end{split}
\end{equation}
We will now add and subtract $B/2\epsilon^i_{\ n}x^n J^0$ to the operators $T^{0i}$ in the equal time commutators and use the $SL(2,\mathbb{R})$ algebra to simplify the expressions
\begin{equation}
\begin{split}
I_{13}(\bb{\epsilon}) &=
\int d^2 \bb{x}\,\, d^2 \bb{\hat{x}} \, e^{-i\bb{\epsilon}\cdot\bb{x}-i\bb{\epsilon}\cdot\bb{\hat{x}}}\left[(\delta'(t-\hat{t})+2\delta(t-\hat{t})\partial_t) \left(\vev{S_{B\,2}^0(t,\bb{x})}\delta^{(2)}(\bb{x}-\hat{\bb{x}})\right.\right.\\
&-\frac{iB}{4}(\bar{\sigma}_1)_{ij}\hat{x}^i \hat{x}^j\vev{[S_{B\,1}^0(t,\bb{x}),J^0(t,\bb{\hat{x}})]}+\frac{iB}{4}(\bar{\sigma}_3)_{ij}x^i x^j\vev{[J^0(t,\bb{x}),S_{B\,3}^0(t,\bb{\hat{x}})]}\\
&\left.+\frac{iB^2}{4}S_{klij}x^ix^j\hat{x}^k\hat{x}^l\vev{[J^0(t,\bb{x}),J^0(t,\bb{\hat{x}})]}\right)\\
&+S_{ijkl}x^i\hat{x}^k\left( \partial_n\partial_{\hat{m}} G_R^{njml}(t-\hat{t};\bb{x},\bb{\hat{x}})-B\epsilon^j_{\ n} \partial_{\hat{m}}G_R^{n,ml}(t-\hat{t};\bb{x},\bb{\hat{x}})\right.\\
&\left.\left.-B\epsilon^l_{\ m} \partial_{n}G_R^{nj,m}(t-\hat{t};\bb{x},\bb{\hat{x}})
+B^2\epsilon^j_{\ n}\epsilon^l_{\ m} G_R^{nm} (t-\hat{t};\bb{x},\bb{\hat{x}})\right) \right].
\end{split}
\end{equation}
We can use the conservation equations of the energy-momentum tensor to rewrite
\begin{equation}
\begin{split}
\partial_{n}G_R^{nj,m}(t-\hat{t};\bb{x},\bb{\hat{x}}) &=-\partial_t G_R^{0j,m}(t-\hat{t};\bb{x},\bb{\hat{x}})+B\epsilon^j_{\ n} G_R^{nm}(t-\hat{t};\bb{x},\bb{\hat{x}})\\
&+i\delta(t-\hat{t})\vev{[T^{0j}(t,\bb{x}),J^m(t,\bb{\hat{x}})]}.
\end{split}
\end{equation}
It will be convenient to write the equal time commutator in the contact term as
\begin{equation}
\begin{split}
\vev{[T^{0j}(t,\bb{x}),J^m(t,\bb{\hat{x}})]} &=\vev{\left[T^{0j}(t,\bb{x})-\frac{B}{2}\epsilon^j_{\ n} x^n J^0(t,\bb{\hat{x}}),J^m(t,\bb{\hat{x}})\right]}\\
&+\frac{B}{2}\epsilon^j_{\ n}x^n\vev{[J^0(t,\bb{x}),J^m(t,\bb{\hat{x}})]}.
\end{split}
\end{equation}
Plugging it back in the expression for $I_{13}$, we find
\begin{equation}
\begin{split}
I_{13}(\bb{\epsilon}) &=
\int d^2 \bb{x}\,\, d^2 \bb{\hat{x}} \, e^{-i\bb{\epsilon}\cdot\bb{x}-i\bb{\epsilon}\cdot\bb{\hat{x}}}\left[(\delta'(t-\hat{t})+2\delta(t-\hat{t})\partial_t) \left(\vev{S_{B\,2}^0(t,\bb{x})}\delta^{(2)}(\bb{x}-\hat{\bb{x}})\right.\right.\\
&-\frac{iB}{4}(\bar{\sigma}_1)_{kl}\hat{x}^k \hat{x}^l\vev{[S_{B\,1}^0(t,\bb{x}),J^0(t,\bb{\hat{x}})]}+\frac{iB}{4}(\bar{\sigma}_3)_{ij}x^i x^j\vev{[J^0(t,\bb{x}),S_{B\,3}^0(t,\bb{\hat{x}})]}\\
&\left.+\frac{iB^2}{4}S_{klij}x^ix^j\hat{x}^k\hat{x}^l\vev{[J^0(t,\bb{x}),J^0(t,\bb{\hat{x}})]}\right)\\
&-\frac{iB}{2}\delta(t-\hat{t})\left((\bar{\sigma}_1)_{kl}\hat{x}^k\vev{[S_{B\,1}(t,\bb{x}),J^l(t,\bb{\hat{x}})]} -(\bar{\sigma}_3)_{ij}x^i\vev{[J^j(t,\bb{x}),S_{B\,3}(t,\bb{\hat{x}})]}\right)\\
&+\frac{iB^2}{2}\delta(t-\hat{t})S_{klij}x^i\hat{x}^k\left(x^j \vev{[J^0(t,\bb{x}),J^l(t,\bb{\hat{x}})]}+\hat{x}^l\vev{[J^j(t,\bb{x}),J^0(t,\bb{\hat{x}})]}\right)\\
&+S_{ijkl}x^i\hat{x}^k\left( \partial_n\partial_{\hat{m}} G_R^{njml}(t-\hat{t};\bb{x},\bb{\hat{x}})+B\epsilon^j_{\ n} \partial_{\hat{t}}G_R^{n,0l}(t-\hat{t};\bb{x},\bb{\hat{x}})\right.\\
&\left.\left.+B\epsilon^l_{\ m} \partial_{t}G_R^{0j,m}(t-\hat{t};\bb{x},\bb{\hat{x}})
-B^2\epsilon^j_{\ n}\epsilon^l_{\ m} G_R^{nm} (t-\hat{t};\bb{x},\bb{\hat{x}})\right) \right].
\end{split}
\end{equation}
The operators $S^0_{B\,a}$  produce infinitesimal deformations
\begin{equation}
i\vev{[S_{B\,a}^0(t,\bb{x}),J^\mu(t,\bb{\hat{x}})]}=-\frac{(\bar{\sigma}_a)_{ij}}{2}x^i\partial_j \vev{J^\mu(\bb{x})}\delta^{(2)}(\bb{x}-\bb{\hat{x}}).
\end{equation}
We show this explicitly for free Dirac fermions in Appendix~\ref{app:shear}.

The equal time commutators of the current should vanish, and time derivatives of expectation values  as well. Then, after integrating by parts and neglecting terms $O(\bb{\epsilon})$
\begin{equation}
\begin{split}
I_{13}(\bb{\epsilon}) &=
\int d^2 \bb{x}\,\, d^2 \bb{\hat{x}} \, e^{-i\bb{\epsilon}\cdot\bb{x}-i\bb{\epsilon}\cdot\bb{\hat{x}}}\left[\delta'(t-\hat{t})\left(\vev{S_{B\,2}^0(\bb{x})}\delta^{(2)}(\bb{x}-\hat{\bb{x}})
-\frac{B}{2}x^2\vev{J^0(\bb{x})}\delta^{(2)}(\bb{x}-\hat{\bb{x}})\right)\right.\\
&-\delta(t-\hat{t})\frac{B}{2}x_i\vev{J^i(\bb{x})}\delta^{(2)}(\bb{x}-\bb{\hat{x}})\\
&+S_{ijkl}x^i\hat{x}^k\left( \partial_n\partial_{\hat{m}} G_R^{njml}(t-\hat{t};\bb{x},\bb{\hat{x}})+B\epsilon^j_{\ n} \partial_{\hat{t}}G_R^{n,0l}(t-\hat{t};\bb{x},\bb{\hat{x}})\right.\\
&\left.\left.+B\epsilon^l_{\ m} \partial_{t}G_R^{0j,m}(t-\hat{t};\bb{x},\bb{\hat{x}})
-B^2\epsilon^j_{\ n}\epsilon^l_{\ m} G_R^{nm} (t-\hat{t};\bb{x},\bb{\hat{x}})\right) \right].
\end{split}
\end{equation}
For a current of the form \eqref{magnetiz}, the contact term proportional to $x_iJ^i$ will drop upon integration by parts. Then, one is left with
\begin{equation}
\begin{split}
I_{13}(\bb{\epsilon}) &=
\int d^2 \bb{x}\, e^{-i2\bb{\epsilon}\cdot\bb{x}}\delta'(t-\hat{t})\left[\vev{S_{B\,2}^0(\bb{x})}-\frac{B}{2}x^2\vev{J^0(\bb{x})}\right]\\
&+\int d^2 \bb{x}\,\, d^2 \bb{\hat{x}} \, e^{-i\bb{\epsilon}\cdot\bb{x}-i\bb{\epsilon}\cdot\bb{\hat{x}}} \,S_{ijkl}x^i\hat{x}^k\left( \partial_n\partial_{\hat{m}} G_R^{njml}(t-\hat{t};\bb{x},\bb{\hat{x}})+B\epsilon^j_{\ n} \partial_{\hat{t}}G_R^{n,0l}(t-\hat{t};\bb{x},\bb{\hat{x}})\right.\\
&\left.+B\epsilon^l_{\ m} \partial_{t}G_R^{0j,m}(t-\hat{t};\bb{x},\bb{\hat{x}})
-B^2\epsilon^j_{\ n}\epsilon^l_{\ m} G_R^{nm} (t-\hat{t};\bb{x},\bb{\hat{x}})\right).
\end{split}
\end{equation}
Using the explicit form of $S_{B\,2}^0$ we get a contribution to the contact term from the angular momentum density and another from the charge density:
\begin{equation}\label{I13magnet}
\begin{split}
I_{13}(\bb{\epsilon}) &=
\int d^2 \bb{x}\, e^{-i2\bb{\epsilon}\cdot\bb{x}}\delta'(t-\hat{t})\left[\vev{S_2^0(\bb{x})}-\frac{B}{4}x^2\vev{J^0(\bb{x})}\right]\\
&+\int d^2 \bb{x}\,\, d^2 \bb{\hat{x}} \, e^{-i\bb{\epsilon}\cdot\bb{x}-i\bb{\epsilon}\cdot\bb{\hat{x}}} \,S_{ijkl}x^i\hat{x}^k\left( \partial_n\partial_{\hat{m}} G_R^{njml}(t-\hat{t};\bb{x},\bb{\hat{x}})+B\epsilon^j_{\ n} \partial_{\hat{t}}G_R^{n,0l}(t-\hat{t};\bb{x},\bb{\hat{x}})\right.\\
&\left.+B\epsilon^l_{\ m} \partial_{t}G_R^{0j,m}(t-\hat{t};\bb{x},\bb{\hat{x}})
-B^2\epsilon^j_{\ n}\epsilon^l_{\ m} G_R^{nm} (t-\hat{t};\bb{x},\bb{\hat{x}})\right) .
\end{split}
\end{equation}
We will now do the Fourier transformation with respect to time
\begin{equation}
\begin{split}
\tilde{I}_{13}(\bb{\epsilon}) &=
\int d^2 \bb{x}\, e^{-i2\bb{\epsilon}\cdot\bb{x}}\,i\omega\left[\vev{S_2^0(\bb{x})}-\frac{B}{4}x^2\vev{J^0(\bb{x})}\right]\\
&+\int d^2 \bb{x}\,\, d^2 \bb{\hat{x}} \, e^{-i\bb{\epsilon}\cdot\bb{x}-i\bb{\epsilon}\cdot\bb{\hat{x}}} \,S_{ijkl}x^i\hat{x}^k\left( \partial_n\partial_{\hat{m}} G_R^{njml}(\omega;\bb{x},\bb{\hat{x}})-i\omega B\epsilon^j_{\ n} G_R^{n,0l}(\omega;\bb{x},\bb{\hat{x}})\right.\\
&\left.+i\omega B\epsilon^l_{\ m} G_R^{0j,m}(\omega;\bb{x},\bb{\hat{x}})
-B^2\epsilon^j_{\ n}\epsilon^l_{\ m} G_R^{nm} (\omega;\bb{x},\bb{\hat{x}})\right).
\end{split}
\end{equation}
The same standard manipulations that we used in the case without magnetic field in the spatial and momentum integrals lead to \eqref{I13solB}.

\subsubsection{Relation to conductivities}
\label{app:mag2}

If rotational invariance is not broken, we can expand the correlation functions as follows (all the form factors $\Pi$ are functions of $\omega$ and $\bb{P^2}$):\footnote{Note that the combination $(P^i\epsilon^{jn}-P^j \epsilon^{in})P_n=-\bb{P^2}\epsilon^{ij}$ is not independent.}
\begin{equation}
\begin{split}\label{gammadecomp}
\Gamma_{AB}^{ij}(\omega,\bb{P})&=\delta^{ij}\Pi_{AB}^{\delta}+\epsilon^{ij}\Pi_{AB}^{\epsilon}+P^i P^j \Pi_{AB}^{p^2}+(P^i \epsilon^{jn} +P^j \epsilon^{in})P_n\Pi_{AB}^{p^2\epsilon},
\end{split}
\end{equation}
where $A,B=J,T$ label current $J^i$ or momentum $T^{0i}$ correlators.

Note that, for any of the correlators with two spatial indices
\begin{equation}
\begin{split}
\frac{\partial^2}{\partial P^i \partial P^k} \Gamma_{AB}^{nm} & = 2\delta^{nm}\delta^{ik}\frac{\partial \Pi_{AB}^\delta}{\partial \bb{P^2}}+2\epsilon^{nm}\delta^{ik}\frac{\partial \Pi_{AB}^\epsilon}{\partial \bb{P^2}}+ (\delta^{nk}\delta^{mi}+\delta^{mk}\delta^{ni})\Pi_{AB}^{p^2}\\
&+(\delta^{nk}\epsilon^{mi}+\delta^{mk}\epsilon^{ni}+\delta^{ni}\epsilon^{mk} +\delta^{mi}\epsilon^{nk} )\Pi_{AB}^{p^2\epsilon}.
\end{split}
\end{equation}
Equating \eqref{I13sol22Delta} to \eqref{I13sol}, we find
\begin{equation}
\eta_H(\omega)=\omega^2\frac{\partial \kappa^\epsilon}{\partial \bb{P^2}}+B^2\frac{\partial \sigma^\epsilon}{\partial \bb{P^2}}+i\omega B\left[\frac{\partial \alpha^\delta}{\partial \bb{P^2}} +\frac{\partial \bar{\alpha}^\delta}{\partial \bb{P^2}}+\frac{1}{2}\left(\alpha^{p^2}+\bar{\alpha}^{p^2}\right)\right] \Bigg|_{\bb{P}=\bb{0}}.
\end{equation}

The first terms are the Hall electric and momentum conductivities defined as
\begin{equation}
\begin{split}
\sigma^\epsilon &= \frac{1}{2}\epsilon_{ij}\sigma^{ij}\equiv \sigma_H,\\
\kappa^\epsilon &= \frac{1}{2}\epsilon_{ij}\kappa^{ij}\equiv \kappa_H.
\end{split}
\end{equation}
The last term depend on the trace of the mixed conductivity
\begin{equation}
\frac{\partial}{\partial \bb{P^2}}\tr\alpha\Bigg|_{\bb{P}=\bb{0}}=2\frac{\partial \alpha^\delta}{\partial \bb{P^2}}+\alpha^{p^2}\Bigg|_{\bb{P}=\bb{0}}.
\end{equation}

\subsection{Coefficients of second derivatives of delta function}

In the presence of magnetic field, terms proportional to the second derivatives of delta function are generated.
These terms cancel out with the following condition
\begin{equation}\label{SecDerB}
\omega^2 \Gamma^{0j0l} (\omega, \bb{0}) = B^2 [\delta^{jl} \delta_{nm} \Gamma^{nm}(\omega, \bb{0}) - \Gamma^{lj}(\omega, \bb{0}) ].
\end{equation}
The relation is strictly valid at zero momentum. It is used to derive (\ref{I13sol22Delta}).

Once we add the drag terms, the terms proportional to the second derivatives of delta function are more complicated.
They need to satisfy
\begin{equation}\label{SecDerBDrag}
\begin{split}
&(\omega^2 + \lambda_T^2) \Gamma^{0j0l} (\omega, \bb{0}) =
(B^j_m - \lambda_J \delta^j_m) (B^l_n - \lambda_J \delta^l_n) \Gamma^{mn}(\omega, \bb{0})\\
&\quad + i(\omega+i \lambda_T) (B^j_m - \lambda_J \delta^j_m)  \Gamma^{m,0l}(\omega, \bb{0}) - i(\omega-i \lambda_T) (B^l_n - \lambda_J \delta^l_n)  \Gamma^{0j,n}(\omega, \bb{0}).
\end{split}
\end{equation}
This relation is used to get (\ref{etahfin}).

\section{Alternative derivation in systems with translation invariance}\label{app:alt}

In this appendix we present an alternative derivation of the Ward identity for the Hall viscosity (at zero magnetic field) in translationally invariant systems that does not require to introduce a regulator.

We will compute the following time derivatives of the $G_{13}$ retarded correlator, integrated over the difference $\mathbf{x}-\hat{\mathbf{x}}$:
\begin{equation}
\begin{split}
&I_{13}\equiv\partial_t\partial_{\hat{t}}\int d^2 (\mathbf{x}-\hat{\mathbf{x}})\,G_{13}(t-\hat{t};\mathbf{x},\hat{\mathbf{x}})=\partial_t\partial_{\hat{t}}\int d^2 (\mathbf{x}-\hat{\mathbf{x}}) S_{ijkl}x^i\hat{x}^k G_R^{0j,0l}(t-\hat{t},\mathbf{x}-\hat{\mathbf{x}})
\end{split}
\end{equation}
We now change variables
\begin{equation}
\mathbf{x}=\mathbf{X}+\half\mathbf{y}, \ \ \mathbf{\mathbf{\hat{x}}}=\mathbf{X}-\half \mathbf{y}.
\end{equation}
Then, we can expand $I_{13}$ in powers of the $X^i$ components as
\begin{equation}
I_{13}(X,t-\hat{t})=X^i X^k A_{ik}(t-\hat{t}) +X^iB_i(t-\hat{t}) +C(t-\hat{t}),
\end{equation}
where, using time-translation invariance
\begin{align}
& A_{ik}= -\frac{\partial^2}{\partial (t-\hat{t})^2}\int d^2 \mathbf{y}\, S_{ijkl} G_R^{0j,0l}(t-\hat{t},\mathbf{y}),\\
& B_i = - \frac{1}{2}\frac{\partial^2}{\partial (t-\hat{t})^2}\int d^2 \mathbf{y}\, \left(S_{kjil}-S_{ijkl}\right) y^k G_R^{0j,0l}(t-\hat{t},\mathbf{y}),\\
&C=\frac{1}{4}\frac{\partial^2}{\partial (t-\hat{t})^2}\int d^2 \mathbf{y}\, S_{ijkl} y^i y^k G_R^{0j,0l}(t-\hat{t},\mathbf{y}).
\end{align}
If we take the time derivatives on $I_{13}$ more explicitly, using the form of the retarded correlator we find
\begin{equation}
\begin{split}
I_{13}&=\int d^2 (\mathbf{x}-\hat{\mathbf{x}}) \left[-i\delta'(t-\hat{t})\vev{\left[S_1^0(t,\mathbf{x}),S_3^0(\hat{t},\hat{\mathbf{x}})\right]}\right.\\
&\left.+i\Theta(t-\hat{t})S_{ijkl}x^i\hat{x}^k\vev{\left[\partial_t T^{0j}(t,\mathbf{x}),\partial_{\hat{t}} T^{0l}(\hat{t},\hat{\mathbf{x}})\right]} \right].
\end{split}
\end{equation}

In the translationally invariant case $\vev{S_2^0(t, \mathbf{x})}=0$.
We will now use translation invariance and the conservation of the energy-momentum tensor to write:
\begin{equation}
\begin{split}
&i\Theta(t-\hat{t})\vev{\left[\partial_t T^{0j}(t,\mathbf{x}),\partial_{\hat{t}} T^{0l}(\hat{t},\hat{\mathbf{x}})\right]} =i\Theta(t-\hat{t})\vev{\left[\partial_n T^{nj}(t,\mathbf{x}),\hat{\partial}_{m} T^{ml}(\hat{t},\hat{\mathbf{x}})\right]}\\
&=\partial_n \hat{\partial}_{m} G_R^{nj,ml}(t-\hat{t},\mathbf{x}-\hat{\mathbf{x}})=-\frac{\partial}{\partial y^n}\frac{\partial}{\partial y^m} G_R^{nj,ml}(t-\hat{t},\mathbf{y}).
\end{split}
\end{equation}
Therefore,
\begin{equation}
\begin{split}
I_{13} &=\int d^2 \mathbf{y}\,\left[-i\delta'(t-\hat{t})\, S_{ijkl}x^i\hat{x}^k \vev{\left[T^{0j}(t,\mathbf{x}),T^{0l}(\hat{t},\hat{\mathbf{x}})\right]}\right.\\
&\left.- S_{ijkl} x^i\hat{x}^k\frac{\partial}{\partial y^n}\frac{\partial}{\partial y^m} G_R^{nj,ml}(t-\hat{t},\mathbf{y})\right].
\end{split}
\end{equation}

As before, we can expand in powers of the $X^i$ components
\begin{equation}
I_{13}(t-\hat{t},\mathbf{X})=X^i X^j \bar{A}_{ij}(t-\hat{t})+X^i\bar{B}_i(t-\hat{t})+\bar{C}(t-\hat{t}),
\end{equation}
where, using translation invariance
\begin{align}
 \bar{A}_{ik}  = & -\int d^2 \mathbf{y}\, S_{ijkl}\left[i\delta'(t-\hat{t})\,  \vev{\left[T^{0j}\left(t,\frac{\mathbf{y}}{2}\right),T^{0l}\left(t,-\frac{\mathbf{y}}{2}\right)\right]}+ \frac{\partial}{\partial y^n}\frac{\partial}{\partial y^m} G_R^{nj,ml}(t-\hat{t},\mathbf{y})\right],\\
\notag \bar{B}_i  =&  -\frac{1}{2}\int d^2 \mathbf{y}\, \left[S_{kjil}-S_{ijkl}\right] y^k \left[ i\delta'(t-\hat{t})\vev{\left[T^{0j}\left(t,\frac{\mathbf{y}}{2}\right),T^{0l}\left(t,-\frac{\mathbf{y}}{2}\right)\right]}\right.\\
&\left.+\frac{\partial}{\partial y^n}\frac{\partial}{\partial y^m} G_R^{nj,ml}(t-\hat{t},\mathbf{y})\right],\\
\notag \bar{C} =&\frac{1}{4}\int d^2 \mathbf{y}\, S_{ijkl} y^i y^k\left[-i\delta'(t-\hat{t})\vev{\left[T^{0j}\left(t,\frac{\mathbf{y}}{2}\right),T^{0l}\left(t,-\frac{\mathbf{y}}{2}\right)\right]}\right.\\
&\left.+ \frac{\partial}{\partial y^n}\frac{\partial}{\partial y^m} G_R^{nj,ml}(t-\hat{t},\mathbf{y})\right].
\end{align}
Since $X^i$ are arbitrary the following conditions must be satisfied $A_{ik}=\bar{A}_{ik}$, $B_i=\bar{B}_i$, $C=\bar{C}$. We now use the equal time commutators. The commutator of $\bar{A}_{ik}$ vanishes because it's the commutator of the momentum densities $T^{0j}$. The commutator in $\bar{B}_i$ is the commutator of the densities $S_{1,3}^0$ and the momentum densities. Assuming the expectation value of the momentum density is zero, this commutator also vanishes. The commutator in $\bar{C}$ is the commutator between $S_1^0$ and $S_3^0$, which is proportional to the angular momentum density:
\begin{equation}
\begin{split}
&\frac{i}{4}\int d^2 \mathbf{y}\, S_{ijkl} y^i y^k\,\delta'(t-\hat{t})\vev{\left[T^{0j}\left(t,\frac{\mathbf{y}}{2}\right),T^{0l}\left(t,-\frac{\mathbf{y}}{2}\right)\right]}\\
&=-i\int d^2 \mathbf{y}\, \,\delta'(t-\hat{t})\vev{\left[S_1^{0}\left(t,\frac{\mathbf{y}}{2}\right),S_3^{0} \left(t,-\frac{\mathbf{y}}{2}\right)\right]}\\
&=-\int d^2 \mathbf{y}\, \,\delta'(t-\hat{t})\vev{S_2^{0}\left(t,\frac{\mathbf{y}}{2}\right)}\delta^{(2)}(\mathbf{y})=-\delta'(t-\hat{t})\vev{S_2^{0}\left(t,\mathbf{0}\right)}=0.
\end{split}
\end{equation}
Then, the condition $C=\bar{C}$ leads to
\begin{equation}
\frac{1}{4}\int d^2 \mathbf{y}\, S_{ijkl} y^i y^k\frac{\partial}{\partial y^n}\frac{\partial}{\partial y^m} G_R^{nj,ml}(t-\hat{t},\mathbf{y}) = \frac{1}{4}\frac{\partial^2}{\partial (t-\hat{t})^2}\int d^2 \mathbf{y}\, S_{ijkl} y^i y^k G_R^{0j,0l}(t-\hat{t},\mathbf{y}).
\end{equation}
We will now do the Fourier transform with respect to time of this expression
\begin{equation}
\int d(t-\hat{t}) e^{-i\omega(t-\hat{t})} \left[C(t-\hat{t})-\bar{C}(t-\hat{t})\right]=0,
\end{equation}
and use the Fourier transform of the retarded correlators
\begin{equation}
G_R^{\mu\nu\alpha\beta}(t-\hat{t},\mathbf{x}-\hat{\mathbf{x}})=\int \frac{dp_0 d^2 \mathbf{p}}{(2\pi)^3} e^{ip_0 (t-\hat{t})+i\mathbf{p}\cdot(\mathbf{x-\hat{x}})}\tilde{G}_R^{\mu\nu\alpha\beta}(p^0,\mathbf{p}).
\end{equation}
We will also use that
\begin{equation}
\begin{split}
&\int d(t-\hat{t}) e^{-i\omega(t-\hat{t})} \int d^2 \mathbf{y}\,  y^i y^k\frac{\partial}{\partial y^n}\frac{\partial}{\partial y^m} G_R^{nj,ml}(t-\hat{t},\mathbf{y})\\
&= \int\frac{d^2 \mathbf{p}}{(2\pi)^2}\int d^2 \mathbf{y}\,\left[\frac{\partial}{\partial p_i}\frac{\partial}{\partial p_k}e^{i\mathbf{p}\cdot\mathbf{y}}\right] p_m p_n \tilde{G}_R^{nj,ml}(\omega,\mathbf{p})\\
&=\int \frac{d^2\mathbf{p}}{(2\pi)^2}\int d^2 \mathbf{y}\frac{\partial}{\partial p_i}\frac{\partial}{\partial p_k}\left[ e^{i\mathbf{p}\cdot\mathbf{y}} p_m p_n \tilde{G}_R^{nj,ml}(\omega,\mathbf{p})\right]\\
&-\frac{\partial}{\partial p_i}\left[e^{i\mathbf{p}\cdot\mathbf{y}}\frac{\partial}{\partial p_k}\left[  p_m p_n \tilde{G}_R^{nj,ml}(\omega,\mathbf{p})\right]\right]+(i\leftrightarrow k)\\
&+e^{i\mathbf{p}\cdot\mathbf{y}}\frac{\partial}{\partial p_i}\frac{\partial}{\partial p_k}\left[  p_m p_n \tilde{G}_R^{nj,ml}(\omega,\mathbf{p})\right].
\end{split}
\end{equation}
We can regulate the momentum integrals with a cutoff $\Lambda$, the first term vanishes since is the derivative of a derivative, while the other derivative terms will vanish upon integration on $\mathbf{y}$, that gives delta functions at zero momentum. Then, the relation becomes
\begin{equation}
 S_{ijkl} \frac{\partial}{\partial p_i}\frac{\partial}{\partial p_k}\left( p_n p_m \tilde{G}_R^{nj,ml}(\omega,\mathbf{p})\right)\Big|_{\mathbf{p}=0} =\omega^2 S_{ijkl}\frac{\partial}{\partial p_i}\frac{\partial}{\partial p_k} \tilde{G}_R^{0j,0l}(\omega,\mathbf{p})\Big|_{\mathbf{p}=0}.
\end{equation}
Which agrees with the result derived in the main text \eqref{ward0}.

\section{Shear generators for free fermions in a magnetic field}
\label{app:shear}

For free Dirac fermions $\psi$ in a background gauge field $A_\mu$ the energy-momentum tensor and current operators are
\begin{equation}
T^\mu_{\ \nu}=-\frac{i}{2}\bar{\psi}\gamma^\mu \overset{\leftrightarrow}{D}_\nu\psi+\frac{1}{2}\delta^\mu_\nu\left(i \bar{\psi} \gamma^\sigma \overset{\leftrightarrow}{D}_\sigma\psi-2m\bar{\psi}\psi\right), \ \ J^\mu =\bar{\psi}\gamma^\mu\psi.
\end{equation}
Where the covariant derivative is $D_\mu \psi=(\partial_\mu -i A_\mu)\psi$ and $D_\mu \bar{\psi}=(\partial_\mu +i A_\mu)\psi$. Using the equations of motion
\begin{equation}
(i\gamma^\mu D_\mu-m) \psi=0 , \ \ iD_\mu\bar{\psi}\gamma^\mu+m\bar{\psi}=0,
\end{equation}
and the algebra of the gamma matrices (the signature of the  metric is mostly minus)
\begin{equation}
 \{\gamma^\mu,\gamma^\nu\}=2\eta^{\mu\nu}\mathbf{1},
\end{equation}
one can check that
\begin{equation}
\partial_\mu T^\mu_{\ \nu}= F_{\nu \sigma} J^\sigma.
\end{equation}
In order to compute the equal time commutators we will use
the following identity for the commutator of composite operators:
\begin{equation}
[AB,CD]=A\{B,C\}D-AC\{B,D\}+\{A,C\}DB-C\{A,D\}B,
\end{equation}
and the equal time anti-commutator of two fermions
\begin{equation}
\begin{split}
&\{ \psi_\alpha(x), \bar{\psi}_\beta(y)\}=\delta^{(2)}(\bb{x}-\bb{y}) \gamma^0_{\alpha\beta},\\
&\{ \psi_\alpha(x), \psi_\beta(y)\}=\{ \bar{\psi}_\alpha(x), \bar{\psi}_\beta(y)\}=0.
\end{split}
\end{equation}
Let us first compute the equal time commutator between two currents
\begin{equation}
\begin{split}
[J^\mu(x),J^\nu(y)]&=\bar{\psi}(x)\left(\gamma^\mu\gamma^0\gamma^\nu -\gamma^\nu\gamma^0\gamma^\mu\right)\psi(x)\, \delta^{(2)}(\bb{x}-\bb{y}).
\end{split}
\end{equation}
If any of the currents is the time component $J^0$ the commutator vanishes, as expected.

The equal time commutator with the momentum density is
\begin{equation}
\begin{split}
&x^i [T^0_{\ j}(x),J^\mu(y)] =i\left(x^i \partial_j J^\mu(x)+\delta^i_j J^\mu(x)\right)\delta^{(2)}(\bb{x}-\bb{y}).
\end{split}
\end{equation}
Then, for
\begin{equation}
S_{B\,a}^0(x)=\frac{(\bar{\sigma}_a)_i^{\ j}}{2}x^i\left[T^0_{\ j}-\frac{B}{2}\epsilon_{j n}x^n J^0 \right].
\end{equation}
The equal time commutator with the current is
\begin{equation}
i\left[ S_{B\,a}^0(x), J^\mu(y)\right]=-\frac{(\bar{\sigma}_a)_{ij}}{2}x^i \partial_j J^\mu(x)\delta^{(2)}(\bb{x}-\bb{y}).
\end{equation}
Note that for the current $S_a^0$ has the same equal time commutator as $S_{B\,a}^0$. However, the action over the fermionic fields is different. Using
\begin{equation}
[AB,C]=A\{B,C\}-\{A,C\}B,
\end{equation}
we find
\begin{equation}
i\left[ S_{B\,a}^0(x), \psi(y)\right]=-\frac{(\bar{\sigma}_a)_{ij}}{2}\left[x^i D_j
-\frac{iB}{2}x^i\epsilon_{jn}x^n \right]\psi(x)\,\delta^{(2)}(\bb{x}-\bb{y}).
\end{equation}
The term proportional to $B$ would be absent in the commutator with $S_a^0$. This term is necessary in order to satisfy the right $SL(2,\mathbb{R})$ algebra. It is most easily seen in the symmetric gauge
\begin{equation}
A_i=-\frac{B}{2}\epsilon_{in}x^n,
\end{equation}
where the commutator reduces to the usual shear transformation
\begin{equation}
i\left[ S_{B\,a}^0(x), \psi(y)\right]=-\frac{(\bar{\sigma}_a)_{ij}}{2}x^i \partial_j
\psi(x)\,\delta^{(2)}(\bb{x}-\bb{y}).
\end{equation}
Here we used the canonical energy-momentum tensor for simplicity, in principle the shear transformations can be generalized for the symmetric energy-momentum tensor.

\section{Spectral decomposition and retarded correlators}
\label{app:spect}

We will label by $\ket{\alpha}$ the eigenstates of the Hamiltonian and denote by $\ket{0}$ the ground state. We can write the correlator of two currents as
\begin{equation}
\begin{split}
\bra{0} V_a^i(t,\bb{x}) V_b^j(\hat{t},\bb{\hat{x}})\ket{0} &=\sum_\alpha \bra{0} V_a^i(t,\bb{x})\ket{\alpha}\bra{\alpha} V_b^j(\hat{t},\bb{\hat{x}})\ket{0} \\
&=\sum_\alpha e^{i\varepsilon_\alpha(t-\hat{t})}\bra{0} V_a^i(\bb{x})\ket{\alpha}\bra{\alpha} V_b^j(\bb{\hat{x}})\ket{0}\\
&=\int \frac{d\omega}{2\pi}e^{i\omega(t-\hat{t})}D_{ab}^{ij}(\omega,\bb{x},\bb{\hat{x}}),
\end{split}
\end{equation}
where $\varepsilon_\alpha$ is the difference between the energies of the state $\ket{\alpha}$ and the ground state. The Fourier transform of the two-point function is then
\begin{equation}
D_{ab}^{ij}=2\pi \sum_\alpha \delta(\omega-\varepsilon_\alpha) \bra{0} V_a^i(\bb{x})\ket{\alpha}\bra{\alpha} V_b^j(\bb{\hat{x}})\ket{0}.
\end{equation}
Using that the step function is ($\epsilon\to 0^+ $)
\begin{equation}
\theta(t-\hat{t})=i\int \frac{d k_0}{2\pi}\frac{e^{-i k_0(t-\hat{t})}}{k_0+i\epsilon},
\end{equation}
the time Fourier transform of the retarded correlator is
\begin{equation}
\begin{split}
\vev{V_a^j V_b^l }_R(\omega,\bb{x},\bb{\hat{x}}) &=\int\frac{d k_0}{2\pi}\frac{1}{\omega-k_0-i\epsilon}\left(D_{ab}^{ij}(k_0,\bb{x},\bb{\hat{x}})-D_{ba}^{ji}(k_0,\bb{\hat{x}},\bb{x}) \right)\\
&=2i\int\frac{d k_0}{2\pi}\frac{1}{\omega-k_0-i\epsilon} \rho_{ab}^{ij}(k_0,\bb{x},\bb{\hat{x}}).
\end{split}
\end{equation}
Where the spectral density is
\begin{equation}
\rho_{ab}^{ij}(k_0,\bb{x},\bb{\hat{x}})=2\pi\sum_\alpha \delta(\omega-\varepsilon_\alpha) \,{\rm Im}\,\left(\bra{0} V_a^i(\bb{x})\ket{\alpha}\bra{\alpha} V_b^j(\bb{\hat{x}})\ket{0}\right).
\end{equation}
Note that for $\ket{\alpha}=\ket{0}$, the expectation value of $V_a^i$ is real, so the ground state contribution drops from the sum.

The Fourier transform respect to space can be defined as
\begin{equation}
\vev{V_a^j V_b^l }_R(\omega,\bb{p},\bb{\hat{q}}) =\int d^2\bb{x}d^2\bb{\hat{x}} e^{-i\bb{p}\cdot{\bb{x}}+i\bb{q}\cdot\bb{\hat{x}}}\vev{V_a^j V_b^l }_R(\omega,\bb{x},\bb{\hat{x}}).
\end{equation}
Then,
\begin{equation}
\begin{split}
\frac{\partial}{\partial p_i}\frac{\partial}{\partial q_k}
 \vev{V_a^j  V_b^l }_R\Bigg|_{\bb{p}=\bb{q}=\bb{0}} &= \int d^2\bb{x}d^2\bb{\hat{x}} x^i\hat{x}^k\vev{V_a^j V_b^l }_R(\omega,\bb{x},\bb{\hat{x}})\\
 &=2i\int\frac{d k_0}{2\pi}\frac{1}{\omega-k_0-i\epsilon} \int d^2\bb{x}d^2\bb{\hat{x}} x^i\hat{x}^k \rho_{ab}^{ij}(k_0,\bb{x},\bb{\hat{x}}).
 \end{split}
\end{equation}

\subsection{Finite temperature}
\label{finiteT}

At finite temperature $T$ the correlators are
\begin{equation}
\begin{split}
\vev{V_a^i(t,\bb{x}) V^j_b(\hat{t},\bb{\hat{x}})}_T &=\tr\left(V_a^i(t,\bb{x}) V^j_b(\hat{t},\bb{\hat{x}}) e^{-(H-E_0)/T} \right)\\
&=
\sum_{\alpha,\beta} e^{-\varepsilon_\beta/T} e^{i(\varepsilon_\alpha-\varepsilon_\beta)(t-\hat{t})}\bra{\beta} V_a^i(\bb{x})\ket{\alpha}\bra{\alpha}V_b^j(\bb{\hat{x}})\ket{\beta}\\
&=
\sum_{\alpha,\beta} e^{-\varepsilon_\alpha/T} e^{-i(\varepsilon_\alpha-\varepsilon_\beta)(t-\hat{t})}\bra{\beta} V_b^j(\bb{\hat{x}})\ket{\alpha}\bra{\alpha}V_a^i(\bb{x})\ket{\beta}.
\end{split}
\end{equation}
Therefore,
\begin{equation}
\begin{split}
D_{ab}^{ij}(\omega,\bb{x},\bb{\hat{x}})_T &= 2\pi\sum_{\alpha,\beta} e^{-\varepsilon_\beta/T}\delta(\omega-(\varepsilon_\alpha-\varepsilon_\beta)) \bra{\beta} V_a^i(\bb{x})\ket{\alpha}\bra{\alpha}V_b^j(\bb{\hat{x}})\ket{\beta}\\
&=\pi\sum_{\alpha,\beta}\left[ e^{-\varepsilon_\beta/T}\delta(\omega-(\varepsilon_\alpha-\varepsilon_\beta)) \bra{\beta} V_a^i(\bb{x})\ket{\alpha}\bra{\alpha}V_b^j(\bb{\hat{x}})\ket{\beta}\right.\\
&\left.+ e^{-\varepsilon_\alpha/T}\delta(\omega+(\varepsilon_\alpha-\varepsilon_\beta)) \bra{\beta}V_b^j(\bb{\hat{x}})\ket{\alpha}\bra{\alpha} V_a^i(\bb{x})\ket{\beta}\right]\\
&=2\pi\delta(\omega)\sum_{\alpha,\beta}\delta_{\varepsilon_\alpha\varepsilon_\beta} e^{-\varepsilon_\beta/T}{\rm Re}\,\left(\bra{\beta} V_a^i(\bb{x})\ket{\alpha}\bra{\alpha}V_b^j(\bb{\hat{x}})\ket{\beta}\right)\\
&+\pi\sum_{\alpha,\beta; \varepsilon_\alpha\neq \varepsilon_\beta}\left[ e^{-\varepsilon_\beta/T}\delta(\omega-(\varepsilon_\alpha-\varepsilon_\beta)) \bra{\beta} V_a^i(\bb{x})\ket{\alpha}\bra{\alpha}V_b^j(\bb{\hat{x}})\ket{\beta}\right.\\
&\left.+ e^{-\varepsilon_\alpha/T}\delta(\omega+(\varepsilon_\alpha-\varepsilon_\beta)) \bra{\beta}V_b^j(\bb{\hat{x}})\ket{\alpha}\bra{\alpha} V_a^i(\bb{x})\ket{\beta}\right].
\end{split}
\end{equation}
The $\delta(\omega)$ term corresponds to the static susceptibilities. The spectral function is then
\begin{equation}
\begin{split}
&\rho_{ab}^{ij}(\omega,\bb{x},\bb{\hat{x}})_T=-\frac{i}{2}\left(D_{ab}^{ij}(\omega,\bb{x},\bb{\hat{x}})_T-D_{ba}^{ji}(\omega,\bb{\hat{x}},\bb{x})_T\right)\\
&=\pi\sum_{\alpha,\beta; \varepsilon_\alpha\neq \varepsilon_\beta}\left[ e^{-\varepsilon_\beta/T}\delta(\omega-(\varepsilon_\alpha-\varepsilon_\beta))\,{\rm Im}\,\left( \bra{\beta} V_a^i(\bb{x})\ket{\alpha}\bra{\alpha}V_b^j(\bb{\hat{x}})\ket{\beta}\right)\right.\\
&\left.+ e^{-\varepsilon_\alpha/T}\delta(\omega+(\varepsilon_\alpha-\varepsilon_\beta)) \,{\rm Im}\,\left(\bra{\beta}V_b^j(\bb{\hat{x}})\ket{\alpha}\bra{\alpha} V_a^i(\bb{x})\ket{\beta}\right)\right].
\end{split}
\end{equation}

The retarded correlator turns out to be
\begin{equation}
\begin{split}
&\vev{V_a^j V_b^l }_R(\omega,\bb{x},\bb{\hat{x}})_T \\
&=i\!\!\!\!\sum_{\alpha,\beta; \varepsilon_\alpha\neq \varepsilon_\beta} \!\!\!\! e^{-\frac{\varepsilon_\beta}{T}}\frac{\omega(1-e^{-\frac{\varepsilon_{\alpha\beta}}{T}}) +\varepsilon_{\alpha\beta} (1+e^{-\frac{\varepsilon_{\alpha\beta}}{T}})}{(\omega-i\epsilon)^2-\varepsilon_{\alpha\beta}^2} ~{\rm Im}\,\left(\bra{\beta} V_a^i(\bb{x})\ket{\alpha}\bra{\alpha} V_b^j(\bb{\hat{x}})\ket{\beta}\right)\\
&=
2i\!\!\!\!\sum_{\alpha, \beta; \varepsilon_\alpha\neq \varepsilon_\beta} \!\!\!\! e^{-\frac{(\varepsilon_\alpha+\varepsilon_\beta)}{2T}}\frac{\omega\sinh\frac{\varepsilon_{\alpha\beta}}{2T} +\varepsilon_{\alpha\beta}\cosh\frac{\varepsilon_{\alpha\beta}}{2T}}{(\omega-i\epsilon)^2-\varepsilon_{\alpha\beta}^2} ~{\rm Im}\,\left(\bra{\beta} V_a^i(\bb{x})\ket{\alpha}\bra{\alpha} V_b^j(\bb{\hat{x}})\ket{\beta}\right),
\end{split}
\end{equation}
where $ \varepsilon_{\alpha\beta} =\varepsilon_\alpha-\varepsilon_\beta$.
The main difference we observe with respect to the zero temperature case is that not only the energy difference with respect to the ground state is relevant, but also the energy differences among all excited states. The matrix elements are suppressed by factors of the frequency in the same way at zero temperature, but there could be a contribution to the Hall viscosity at non-zero magnetic field or if the density of states grows at small frequencies.



\begin{thebibliography}{999}

\bibitem{Avron:1995}
{{Avron}, J.~E. and {Seiler}, R. and {Zograf}, P.~G.},
"{Viscosity of Quantum Hall Fluids}",
{Phys.~Rev~ Lett.} {\bf 75}, 697 (1995).
arXiv: cond-mat/9502011


\bibitem{Landau10}
\newblock E.~Lifshitz and L.~Pitaevskii, {\em Landau and Lifshitz, Course in
  Theoretical Physics} Vol. 10: Physical Kinetics (Pergamon Press, Oxford,
  1981).

\bibitem{Avron:1997}
J.~E. {Avron},
"{Odd Viscosity}"
 arXiv:physics/9712050.

\bibitem{Hoyos:2014pba}
  C.~Hoyos,
  ``Hall viscosity, topological states and effective theories,''
  Int.\ J.\ Mod.\ Phys.\ B {\bf 28}, 1430007 (2014)
  [arXiv:1403.4739 [cond-mat.mes-hall]].

\bibitem{Levay1995}
P.~L\'evay,
``Berry phases for Landau Hamiltonians on deformed tori.''
J. Math. Phys. {\bf 36}, 2792 (1995).


\bibitem{Tokatly2007}
I.~V. {Tokatly} and G.~{Vignale},
``Lorentz shear modulus of a two-dimensional electron gas at high magnetic field.''
\newblock Phys. Rev. B {\bf 76}, 161305 (2007),
Phys. Rev. B {\bf 79}, 199903 (2009).
arXiv: 0706.2454.

\bibitem{Read:2008rn}
  N.~Read,
  ``Non-Abelian adiabatic statistics and Hall viscosity in quantum Hall states and p(x) + ip(y) paired superfluids,''
  Phys.\ Rev.\ B {\bf 79}, 045308 (2009)
  [arXiv:0805.2507 [cond-mat.mes-hall]].

\bibitem{Tokatly2009}
I.~V. {Tokatly} and G.~{Vignale},
``Lorentz shear modulus of fractional quantum Hall states.''
Journal of Physics Condensed Matter {\bf 21}, A265603 (2009),
  0812.4331.

\bibitem{Read2011}
N.~{Read} and E.~H. {Rezayi},
``Hall viscosity, orbital spin, and geometry: paired superfluids and quantum Hall systems.''
Phys. Rev. B {\bf 84}, 085316 (2011), 1008.0210.
arXiv:1008.0210

\bibitem{Bradlyn:2012ea}
  B.~Bradlyn, M.~Goldstein and N.~Read,
  ``Kubo formulas for viscosity: Hall viscosity, Ward identities, and the relation with conductivity,''
  Phys.\ Rev.\ B {\bf 86}, 245309 (2012)
  [arXiv:1207.7021 [cond-mat.stat-mech]].


\bibitem{Cho2014}
G.~Y.~Cho, Y.~You and E.~Fradkin,
``Field Theory of the Geometry of Fractional Quantum Hall Fluids,''
[arXiv: 1406.2700 [cond-mat.str-el]]

\bibitem{Hughes:2011hv}
  T.~L.~Hughes, R.~G.~Leigh and E.~Fradkin,
  ``Torsional Response and Dissipationless Viscosity in Topological Insulators,''
  Phys.\ Rev.\ Lett.\  {\bf 107}, 075502 (2011)
  [arXiv:1101.3541 [cond-mat.mes-hall]].

\bibitem{Hughes:2012vg}
  T.~L.~Hughes, R.~G.~Leigh and O.~Parrikar,
  ``Torsional Anomalies, Hall Viscosity, and Bulk-boundary Correspondence in Topological States,''
  Phys.\ Rev.\ D {\bf 88}, no. 2, 025040 (2013)
  [arXiv:1211.6442 [hep-th]].

\bibitem{Nicolis:2011ey}
  A.~Nicolis and D.~T.~Son,
  ``Hall viscosity from effective field theory,''
  arXiv:1103.2137 [hep-th].

\bibitem{Hoyos:2011ez}
  C.~Hoyos and D.~T.~Son,
  ``Hall Viscosity and Electromagnetic Response,''
  Phys.\ Rev.\ Lett.\  {\bf 108}, 066805 (2012)
  [arXiv:1109.2651 [cond-mat.mes-hall]].

\bibitem{Hoyos:2013eha}
  C.~Hoyos, S.~Moroz and D.~T.~Son,
  ``Effective theory of chiral two-dimensional superfluids,''
  Phys.\ Rev.\ B {\bf 89}, 174507 (2014)
  [arXiv:1305.3925 [cond-mat.quant-gas]].

\bibitem{Haehl:2013kra}
  F.~M.~Haehl and M.~Rangamani,
  ``Comments on Hall transport from effective actions,''
  JHEP {\bf 1310}, 074 (2013)
  [arXiv:1305.6968 [hep-th]].

\bibitem{Geracie:2014iva}
  M.~Geracie and D.~T.~Son,
  ``Effective field theory for fluids: Hall viscosity and Wess-Zumino-Witten term,''
  arXiv:1402.1146 [hep-th].

\bibitem{Hidaka:2013}
Y.~Hidaka, Y.~Hirono, T.~Kimura and Y.~Minami,
``Viscoelastic-electromagnetism and Hall viscosity,''
  Progress of Theoretical and Experimental Physics, no. 1, 010003 (2013)
[arXiv:1206.0734 [cond-mat.mes-hall]].

\bibitem{Jensen:2011xb}
  K.~Jensen, M.~Kaminski, P.~Kovtun, R.~Meyer, A.~Ritz and A.~Yarom,
  ``Parity-Violating Hydrodynamics in 2+1 Dimensions,''
  JHEP {\bf 1205}, 102 (2012)
  [arXiv:1112.4498 [hep-th]].


\bibitem{Jensen:2012jh}
  K.~Jensen, M.~Kaminski, P.~Kovtun, R.~Meyer, A.~Ritz and A.~Yarom,
  ``Towards hydrodynamics without an entropy current,''
  Phys.\ Rev.\ Lett.\  {\bf 109}, 101601 (2012)
  [arXiv:1203.3556 [hep-th]].

\bibitem{Banerjee:2012iz}
  N.~Banerjee, J.~Bhattacharya, S.~Bhattacharyya, S.~Jain, S.~Minwalla and T.~Sharma,
  ``Constraints on Fluid Dynamics from Equilibrium Partition Functions,''
  JHEP {\bf 1209}, 046 (2012)
  [arXiv:1203.3544 [hep-th]].




\bibitem{Kaminski2013}
M.~{Kaminski} and S.~{Moroz},
``Non-Relativistic Parity-Violating Hydrodynamics in Two Spatial Dimensions.''
\newblock ArXiv e-prints  (2013), 1310.8305.

\bibitem{Son:2013xra}
  D.~T.~Son and C.~Wu,
 ``Holographic Spontaneous Parity Breaking and Emergent Hall Viscosity and Angular Momentum,''
  arXiv:1311.4882 [hep-th].

\bibitem{Saremi:2011ab}
O.~Saremi and D.~T. Son,
``Hall viscosity from gauge/gravity duality.''
\newblock JHEP {\bf 1204}, 091 (2012), 1103.4851.

\bibitem{Liu:2014gto}
  H.~Liu, H.~Ooguri and B.~Stoica,
  ``Hall Viscosity and Angular Momentum in Gapless Holographic Models,''
  arXiv:1403.6047 [hep-th].

\bibitem{Hoyos:2014nua}
  C.~Hoyos, B.~S.~Kim and Y.~Oz,
  ``Odd Parity Transport In Non-Abelian Superfluids From Symmetry Locking,''
  arXiv:1404.7507 [hep-th].

\bibitem{Chen2011}
J.-W. Chen, N.-E. Lee, D.~Maity, and W.-Y. Wen,
``A Holographic Model For Hall Viscosity.''
\newblock Phys.Lett. {\bf B713}, 47 (2012), 1110.0793.

\bibitem{Chen2012}
J.-W. Chen, S.-H. Dai, N.-E. Lee, and D.~Maity,
``Novel Parity Violating Transport Coefficients in 2+1 Dimensions from Holography.''
\newblock JHEP {\bf 1209}, 096 (2012), 1206.0850.

\bibitem{Cai:2012mg}
R.-G. Cai, T.-J. Li, Y.-H. Qi, and Y.-L. Zhang,
``Incompressible Navier-Stokes Equations from Einstein Gravity with Chern-Simons Term.''
\newblock Phys.Rev. {\bf D86}, 086008 (2012), 1208.0658.

\bibitem{Wu:2013vya}
  C.~Wu,
  ``Angular Momentum Generation from Holographic Chern-Simons Models,''
  arXiv:1311.6368 [hep-th].

\bibitem{Liu:2012zm}
  H.~Liu, H.~Ooguri, B.~Stoica and N.~Yunes,
  ``Spontaneous Generation of Angular Momentum in Holographic Theories,''
  Phys.\ Rev.\ Lett.\  {\bf 110}, no. 21, 211601 (2013)
  [arXiv:1212.3666 [hep-th]].  
  \\   H.~Liu, H.~Ooguri and B.~Stoica,
  ``Angular Momentum Generation by Parity Violation,''
  arXiv:1311.5879 [hep-th].

\bibitem{Mendoza2013}
M.~Mendoza, H.J.~Herrmann, S.~Succi,
``Hydrodynamic Model for Conductivity in Graphene,''
Scientific Reports {\bf 3},1052 (2013)
[arXiv:1301.3428 [cond-mat.mes-hall]]

\bibitem{Hartnoll2007}
S.A.~Hartnoll,  P.K.~Kovtun, M.~M{\"u}ller, S.~Sachdev,
``Theory of the Nernst effect near quantum phase transitions in condensed matter and in dyonic black holes,''
Phys.\ Rev.\ B\ {\bf 76},144502 (2007)
[arXiv:0706.3215 [cond-mat.str-el]]


\bibitem{Geracie:2014}
M.~Geracie, D.~Thanh Son, C.~Wu, S.-F.~Wu,
``Spacetime Symmetries of the Quantum Hall Effect,''
  [arXiv: 1407.1252 [cond-mat.mes-hall]]

\bibitem{Gromov2014a}
A.~Gromov, A.G.~Abanov,
``Density-curvature response and gravitational anomaly,''
[arXiv: 1403.5809 [cond-mat.str-el]]

\bibitem{Gromov2014}
A.~Gromov, A.~G.~Abanov,
``Thermal Hall Effect and Geometry with Torsion,''
[arXiv:1407.2908 [cond-mat.str-el]]


\bibitem{Bradlyn2014}
B.~Bradlyn and N.~Read,
``Low-energy effective theory in the bulk for transport in a topological phase,''
[arXiv:1407.2911 [cond-mat.mes-hall]]

\bibitem{Gubser:2008zu}
  S.~S.~Gubser,
  ``Colorful horizons with charge in anti-de Sitter space,''
  Phys.\ Rev.\ Lett.\  {\bf 101}, 191601 (2008)
  [arXiv:0803.3483 [hep-th]].

\bibitem{Ardonne:2003}
E.~Ardonne, P.~Fendley, E.~Fradkin,
``Topological order and conformal quantum critical points,''
Annals of Physics, {\bf 310},  493-551 (2004)
[cond-mat/0311466]

\bibitem{Kachru:2008yh}
  S.~Kachru, X.~Liu and M.~Mulligan,
  ``Gravity duals of Lifshitz-like fixed points,''
  Phys.\ Rev.\ D {\bf 78}, 106005 (2008)
  [arXiv:0808.1725 [hep-th]].

\bibitem{Hoyos:2013eza}
  C.~Hoyos, B.~S.~Kim and Y.~Oz,
  ``Lifshitz Hydrodynamics,''
  JHEP {\bf 1311}, 145 (2013)
  [arXiv:1304.7481 [hep-th]].
  C.~Hoyos, B.~S.~Kim and Y.~Oz,
  ``Lifshitz Field Theories at Non-Zero Temperature, Hydrodynamics and Gravity,''
  JHEP {\bf 1403}, 029 (2014)
  [arXiv:1309.6794 [hep-th]].




\end{thebibliography}
\end{document}